\documentclass[aps,pra,twocolumn,longbibliography,superscriptaddress,floatfix]{revtex4-1}
\usepackage{amsfonts}
\usepackage{graphicx}
\usepackage{epsfig}
\usepackage{dcolumn}
\usepackage{bm}
\usepackage{amsmath}
\usepackage{graphicx}
\usepackage[latin1]{inputenc}
\usepackage{ulem}
\usepackage{epstopdf}
\usepackage{subfigure}
\usepackage{color}
\usepackage{amsthm}
\usepackage{newlfont}
\usepackage{graphicx}
\usepackage{epstopdf}
\usepackage{appendix}
\usepackage[breaklinks=true]{hyperref}
\usepackage{breakcites}
\usepackage{textcomp}
\usepackage{appendix}
\usepackage{multirow}
\usepackage{color}
\usepackage{amssymb}
\usepackage{epsfig}
\usepackage{mathptmx}
\usepackage{bm}
\usepackage[american]{babel}
\usepackage{braket}

\begin{document}

\title{Modulational instability, inter-component asymmetry
and formation of quantum droplets in one-dimensional binary Bose gases}
\author{Thudiyangal Mithun}
\affiliation{Center for Theoretical Physics of Complex Systems, Institute for Basic
Science, Daejeon, Korea}
\affiliation{Department of Mathematics and Statistics, University of Massachusetts, Amherst MA 01003-4515, USA}
\author{Aleksandra Maluckov}
\affiliation{Vinca Institute of Nuclear Sciences, University of Belgrade, P.
O. B. 522,11001 Belgrade, Serbia}
\author{Kenichi Kasamatsu}
\affiliation{Department of Physics, Kindai University, Higashi-Osaka, Osaka
577-8502, Japan}
\author{Boris A. Malomed}
\affiliation{Department of Physical Electronics, School of Electrical Engineering,
Faculty of Engineering, and Center for Light-Matter Interaction, Tel Aviv University, Tel Aviv 69978, Israel}
\author{Avinash Khare}
\affiliation{Department of Physics, Savitribai Phule Pune University, Pune
411007, India}

\begin{abstract}
Quantum droplets are ultradilute liquid states which emerge from the
competitive interplay of two Hamiltonian terms, the mean-field energy
and beyond-mean-field correction, in a weakly interacting binary Bose gas. We relate the formation of
droplets in symmetric and asymmetric  two-component one-dimensional boson systems to the
modulational instability of a spatially uniform state driven by the beyond-mean-field term. Asymmetry between the components may be
caused by their unequal populations or unequal intra-component interaction strengths. Stability of both symmetric and asymmetric droplets is investigated. Robustness of the symmetric solutions against symmetry-breaking perturbations is confirmed.
\end{abstract}

\maketitle


\section{Introduction}

The mean-field (MF) theory of weakly interacting dilute atomic gases rules
out formation of a liquid state \cite{pitaevskii2016bose,Pethick2002}.
However, it has been recently shown that a liquid phase arises if one takes
into account beyond-MF effects originating from quantum fluctuations around
the MF ground state of weakly interacting binary (two-component) Bose gases
\cite{Petrov:2015}. A fundamental property which allows one to interpret
this phase as a fluid is incompressibility: it maintains a limit density
which cannot be made larger (see details below), hence adding more atoms
leads to spatial expansion of the state. Another fundamental feature of this
quantum-fluid phase is that it facilitates self-trapping of \textit{quantum
droplets} (QDs), which are stabilized by the interplay between the contact
MF interaction and the beyond-MF Lee-Huang-Yang (LHY) correction \cite%
{LHY1957}. Binary Bose-Einstein condensates (BECs) with competing intra- and
inter component MF interactions of opposite signs offer a remarkable
possibility for the generation of QDs, as proposed by Petrov \cite%
{Petrov:2015}. This possibility was further elaborated in various settings,
including different effective dimensions \cite%
{Petrov:2016,Li:2017,Luca1,Jorgensen:2018Dilute,cikojevic2018ultradilute,Luca2,Kartashov:2018Three,Astrakharchik:2018Dynamics,crossover,Li:2018Two,Ancilotto2018,Liu2019,Chiquillo2019,Tononi2019soc,Kartashov2019,semi-discrete}%
. In particular, the dynamics of QDs with the flat-top (FT) or Gaussian
shape, which correspond to large or relatively small numbers of particles,
respectively, was addressed in the framework of the one-dimensional (1D)
reduction of the model \cite{Astrakharchik:2018Dynamics}. The theoretical
prediction was followed by experimental creation of QDs in mixtures of two
different atomic states of $^{39}$K, with quasi-2D \cite%
{cabrera2018quantum,Cheiney:2018Bright} and fully 3D \cite%
{Semeghini:2018Self,2018arXiv181209151F} shapes (see also recent reviews
\cite{review,todayrev}). Very recently, the creation of especially
long-lived QDs was reported in a heteronuclear $^{41}$K-$^{87}$Rb system
\cite{long-lived}. Another theoretically predicted and experimentally
realized option for the creation of QDs makes use of the single-component
condensate with dipole-dipole interactions \cite%
{Rosenzweig,Pfau,Wachtler:2016Quantum,Ferrier:2016Observation,Wachtler:2016Ground,Baillie:2018Droplet,Ferrier-BarbutOnset,no-stable-vortex}%
. It is relevant to mention that formation of multiple droplets was also
predicted and experimentally observed as an MF effect in strongly
nonequilibrium (turbulent) states of BECs \cite{Yuk}.

Collective modes of QDs are a subject of special interest, as they reveal
internal dynamics of the droplets \cite%
{Bulgac:2002Dilute,Wachtler:2016Ground,Baillie:2017Collective,Astrakharchik:2018Dynamics,2018arXiv181209151F}%
. In particular, the stable existence of the QDs is secured if the
particle-emission threshold lies below all excitation modes, hence a
perturbation in the form of such modes will not cause decay of the droplet.

We here aim to address issues which are related to the creation of QDs in
the 1D setting and were not addressed in previous works. First, we consider
modulational instability (MI) of spatially uniform plane-wave (PW) states,
in the framework of the coupled system of Gross-Pitaevskii (GP) equations
with the LHY corrections, for the two-component MF wave function of the
binary condensate. This is the system which was originally derived in Ref.
\cite{Petrov:2016}. Recently, MI has been experimentally demonstrated in
BECs with attractive interactions \cite%
{Nguyen:2017Formation,Everitt:2017Observation,SanzInteraction}. Other
examples of the MI are provided by the binary BEC with the linear Rabi
coupling or the spin-orbit coupling \cite{SanzInteraction,Ponz}, and by a
system combining the MF and LHY terms \cite{Singh}. The linear-stability
analysis, followed by direct simulations of the corresponding GP equations,
shows that the lower branch of the PW states exhibits MI, the instability
splitting the PW into a chain of localized droplet-like structures.
Secondly, we address properties of the QDs in the binary condensate in the
framework of the two-component GP system, without assuming effective
inter-component symmetry, which reduces the system to a single-component GP
equation. The asymmetry implies different MF self-repulsion coefficients in
the two components, and/or unequal norms in them. Although properties of QDs
have been studied by using the two-component GP system in some papers \cite%
{Li:2017,Kartashov:2018Three,Ancilotto2018,Liu2019,Tononi2019soc,Kartashov2019}%
, the explicit asymmetry of the system parameters has not been addressed,
except for Ref.~\cite{Ancilotto2018} in which the situation for $^{39}$K-$%
^{39}$K and $^{23}$Na-$^{87}$Rb atomic mixtures have been considered. We
conclude that the population difference between the components does not
significantly affect density profiles of QDs in the system with equal MF
self-repulsion strengths in the two components. On the other hand,we find
that profiles of the QD solutions are essentially asymmetric when the
self-repulsion coefficients are different in the components. Generally, the
numerical findings corroborate stability of the known symmetric states
against symmetry-breaking perturbations. We also address the MI of the
two-component system, and demonstrate that chains of asymmetric QDs can be
generated by the MI-induced nonlinear evolution.

The paper is organized as follows. In Sec.~\ref{sec1} we introduce the model
and discuss conditions necessary for the formation of the droplets. Section~%
\ref{sec2} deals with the single-component version of the symmetric system.
We consider various solutions admitted by it (PW, FT, periodic, etc.), and
apply the linear-stability analysis of the PW solution to assess the MI, in
a combination with direct simulations. In Sec.~\ref{sec3}, we address the
stability of asymmetric droplets, as well as the formation of droplets in
the two-component asymmetric system via the MI. The paper is concluded by
Sec.~\ref{sec5}. Additional symmetric and asymmetric exact and approximate
analytical solutions are presented in Appendices.

\section{Model and methods}

{\label{sec1}} We consider the 1D model of the two-component condensate with
coefficients of the intra-component repulsion, $g_{1}>0$ and $g_{2}>0$, and
inter-component attraction, $g_{12}<0$. In the weak-interaction limit, the
corresponding energy density, which includes the MF terms and LHY
correction, was derived in Ref. \cite{Petrov:2016}:

\begin{equation}
\begin{split}
\mathcal{E}_{\mathrm{1D}}&=\frac{\left( \sqrt{g_{1}}\rho _{1}-\sqrt{g_{2}}%
\rho _{2}\right) ^{2}}{2}+\frac{g\delta g\left( \sqrt{g_{2}}\rho _{1}+\sqrt{%
g_{1}}\rho _{2}\right) ^{2}}{(g_{1}+g_{2})^{2}}\\&-\frac{2\sqrt{m}\left(
g_{1}\rho _{1}+g_{2}\rho _{2}\right) ^{3/2}}{3\pi \hbar },  \label{E1D}
\end{split}
\end{equation}%
where $m$ is the atomic mass (the same for both components), $\rho
_{j}=|\Psi _{j}|^{2}$ $(j=1,2)$ is the density of the $j$-th component,
represented by the MF wave function $\Psi _{j}$, and

\begin{equation}
g\equiv \sqrt{g_{1}g_{2}},\quad \quad \delta g\equiv g_{12}+g.  \label{g}
\end{equation}%
The last term in Eq.~(\ref{E1D}) represents the LHY correction. Derivation
of Eq.~\eqref{E1D} assumes that the binary BEC is close to the point of the
MF repulsion-attraction balance, with $\left\vert \delta g\right\vert \ll g$
. In experiments, $\delta g$ may be tuned to be both positive and negative
\cite{cabrera2018quantum,Cheiney:2018Bright,Semeghini:2018Self}.

Equation~\eqref{E1D} is valid in the case of tight confinement applied in
the transverse dimensions, which makes the setting effectively
one-dimensional. In the 3D case, the LHY term $\sim -\rho ^{3/2}$ (for $\rho
_{1}=\rho _{2}\equiv \rho $) is replaced by one $\sim +\rho ^{5/2}$. A
detailed consideration of the crossover from 3D to 1D \cite%
{crossover,Ilg:2018,Edler:2018} in the two-component system is a problem
which may be a subject of a separate work. Here, it is relevant to compare
the symmetric version of Eq.~(\ref{E1D}) for the energy density with that
recently presented in Ref. \cite{crossover}. It demonstrates that an
accurately derived LHY contribution to the energy density of the 1D system
contains, in addition to the $\rho ^{3/2}$ term which was derived in Ref.~%
\cite{Petrov:2016}, a term $\sim \rho ^{2}$, which can be absorbed into the
mean-field energy density, and a higher-order term $\sim \rho ^{3}$, which
was omitted in the analysis reported in Ref.~\cite{crossover}. A conclusion
formulated in that work is that the energy density originally derived in
Ref.~\cite{Petrov:2016} is literally valid if the ratio of the mean-field
energy to that of the transverse confinement takes values $\leq 0.03$. For
typical experimental parameters, this implies that the difference between
absolute values of scattering lengths of the mean-field intra-component
repulsion and inter-component attraction should be $\leq 1$ nm, which may be
achieved in the experiment. The 1D QDs originate from the balance of the
second term in Eq.~(\ref{E1D}), corresponding to the weakly repulsive MF
interaction, with $\delta g>0$, and the LHY term, which introduces effective
attraction in the 1D setting, on the contrary to the repulsion in the 3D
setting \cite{Petrov:2016,Astrakharchik:2018Dynamics}.

The energy functional, $\int_{-\infty }^{+\infty }\mathcal{E}_{\text{1D}}dZ$%
, gives rise to the system of GP equations, which include the LHY correction,

\begin{equation}
\begin{split}
i\hbar \frac{\partial \Psi _{1}}{\partial T}& =-\frac{\hbar ^{2}}{2m}\frac{%
\partial ^{2}\Psi _{1}}{\partial Z^{2}}+(g_{1}+Gg_{2})|\Psi _{1}|^{2}\Psi
_{1}-(1-G)g|\Psi _{2}|^{2}\Psi _{1}\\&-\frac{g_{1}\sqrt{m}}{\pi \hbar }\sqrt{%
g_{1}|\Psi _{1}|^{2}+g_{2}|\Psi _{2}|^{2}}\Psi _{1}, \\
i\hbar \frac{\partial \Psi _{2}}{\partial T}& =-\frac{\hbar ^{2}}{2m}\frac{%
\partial ^{2}\Psi _{2}}{\partial Z^{2}}+(g_{2}+Gg_{1})|\Psi _{2}|^{2}\Psi
_{2}-(1-G)g|\Psi _{1}|^{2}\Psi _{2}\\&-\frac{g_{2}\sqrt{m}}{\pi \hbar }\sqrt{%
g_{1}|\Psi _{1}|^{2}+g_{2}|\Psi _{2}|^{2}}\Psi _{2},
\end{split}
\label{Psi}
\end{equation}%
where $T$ and $Z$ are the time and coordinate measured in physical units,
and parameter
\begin{equation}
G=\frac{2g\delta g}{(g_{1}+g_{2})^{2}},  \label{G}
\end{equation}%
measures the deviation from the MF repulsion-attraction balance point, see
Eq. (\ref{g}). The normalization of the components of the wave function is
determined by numbers of bosons in each component:

\begin{equation}
N_{j}=\int_{-\infty }^{+\infty }|\Psi _{j}|^{2}dZ.  \label{N}
\end{equation}

Further, rescaling

\begin{equation}
\left( \frac{mg^{2}}{\hbar ^{3}}\right) T\equiv t,\quad \left( \frac{mg}{%
\hbar ^{2}}\right) Z\equiv z,\quad \left( \frac{\hbar }{\sqrt{mg}}\right)
\Psi _{1,2}\equiv \psi _{1,2}  \label{rescale}
\end{equation}%
casts Eq.~(\ref{Psi}) in the normalized form,

\begin{equation}
\begin{split}
i\frac{\partial \psi _{1}}{\partial t}& =-\frac{1}{2}\frac{\partial ^{2}\psi
_{1}}{\partial z^{2}}+(P+GP^{-1})|\psi _{1}|^{2}\psi _{1}-(1-G)|\psi
_{2}|^{2}\psi _{1}\\&-\frac{P}{\pi }\sqrt{P|\psi _{1}|^{2}+P^{-1}|\psi _{2}|^{2}%
}\psi _{1}, \\
i\frac{\partial \psi _{2}}{\partial t}& =-\frac{1}{2}\frac{\partial ^{2}\psi
_{2}}{\partial z^{2}}+(P^{-1}+GP)|\psi _{2}|^{2}\psi _{2}-(1-G)|\psi
_{1}|^{2}\psi _{2}\\&-\frac{1}{\pi P}\sqrt{P^{-1}|\psi _{2}|^{2}
+P|\psi _{1}|^{2}}\psi _{2},
\end{split}
\label{2comp_GP}
\end{equation}%
where parameter

\begin{equation}
P\equiv \sqrt{\frac{g_{1}}{g_{2}}}=\frac{g_{1}}{g}  \label{PG}
\end{equation}%
determines the asymmetry of the system, in the case of $P\neq 1$. Note that,
as concerns stationary solutions with chemical potentials $\mu _{1,2}$,
sought for as

\begin{equation}
\psi _{1,2}\left( z,t\right) =\exp (-i\mu _{1,2}t)\phi _{1,2}(z),  \label{mu}
\end{equation}%
states with mutually proportional components, $\phi _{1}(z)=K\phi _{2}(z)$,
are only possible in the fully symmetric case with $P=1$, $\mu _{1}=\mu _{2}$%
, and $K=1$. In previous works \cite{Petrov:2016,Astrakharchik:2018Dynamics}%
, 1D solutions for QDs were considered only in the framework of the single
GP equation which corresponds to symmetric system \eqref{2comp_GP} with $%
P=1 $ and $\psi _{1}=\psi _{2}$.


\section{Modulation instability versus QDs}

In this section we address MI\ of PWs in both symmetric and asymmetric GP
systems, and relate it to formation of the QDs in the binary bosonic gas. To
the best of our knowledge, this is the first work aiming to associate the MI
with the formation of the 1D droplets in the system with unequal components.
We first consider MI in the framework of the single-component reduction of
the symmetric version of system (\ref{2comp_GP}), after briefly reviewing
stationary solutions of the GP equation. Next, we extend the analysis for
the two-component GP system, which makes it possible to produce asymmetric
QDs, starting from the MI of asymmetric PW states.

\subsection{The single-component GP model}

{\label{sec2}} Under the single-component reduction of the binary system,
with $g_{1}=g_{2}\equiv g$ and $\psi _{1}=\psi _{2}\equiv \psi $, Eq.~(\ref%
{E1D}) simplifies to \cite{Petrov:2016}

\begin{equation}
\varepsilon _{\mathrm{1D}}\equiv \frac{\hbar ^{4}}{m^{2}g^{3}}\mathcal{E}_{%
\mathrm{1D}}=\frac{\delta g}{g}n^{2}-\frac{2^{5/2}}{3\pi }n^{3/2},
\label{Esymm}
\end{equation}%
with the single dimensionless density, $n=|\psi |^{2}\equiv \left( \hbar
^{2}/mg\right) \rho $. Assuming a spatially uniform state, the equilibrium
density and the corresponding chemical potential are given by

\begin{equation}
n_{0}=\frac{8}{9\pi ^{2}}\left( \frac{g}{\delta g}\right) ^{2},~\mu _{0}=-%
\frac{4}{9\pi ^{2}}\frac{g}{\delta g}.  \label{1n0}
\end{equation}%
Density $n_{0}$ corresponds to the minimum of the energy per particle, $%
\partial _{n}\left[ n^{-1}\varepsilon _{\mathrm{1D}}(n)\right] =0$, and $\mu
_{0}$ is negative for $\delta g/g>0$. The corresponding single GP equation is

\begin{equation}
i\frac{\partial \psi }{\partial t}=-\frac{1}{2}\frac{\partial ^{2}\psi }{%
\partial z^{2}}+\frac{\delta g}{g}|\psi |^{2}\psi -\frac{\sqrt{2}}{\pi }%
|\psi |\psi ,  \label{eq13}
\end{equation}%
with normalization condition $\int_{-\infty }^{+\infty }|\psi (z)|^{2}dz=N$,
where $N\equiv N_{1}=N_{2}$ is the number of atoms in each component.

Although coefficient $\delta g/g$ can be scaled out in Eq.~(\ref{eq13}), as
done in Ref.~\cite{Astrakharchik:2018Dynamics}, we keep it here as a free
parameter. This option is convenient for the subsequent consideration of the
MI, treating $\delta g/g$ and density $n$ as independent constants, which
may be matched to experimentally relevant parameters.

Below, we address two stationary solutions of Eq.~\eqref{eq13}. One is the
QD bound state of a finite size, which was studied in detail in Refs.~\cite%
{Petrov:2016} and \cite{Astrakharchik:2018Dynamics}. The other solution is
the PW with uniform density. Here we briefly recapitulated basic properties
of these solutions for the completeness of the presentation. In subsection~%
\ref{MIsingled} we address MI of the PWs and associate it with the
spontaneous generation of chains of localized modes. Additional families of
exact analytical solutions of Eq.~\eqref{eq13} are given in Appendix \ref%
{othersol1}.

\subsubsection{The droplet solution}

\label{flattops} 
\begin{figure}[tbh]
\centering
\includegraphics[width=0.95\columnwidth]{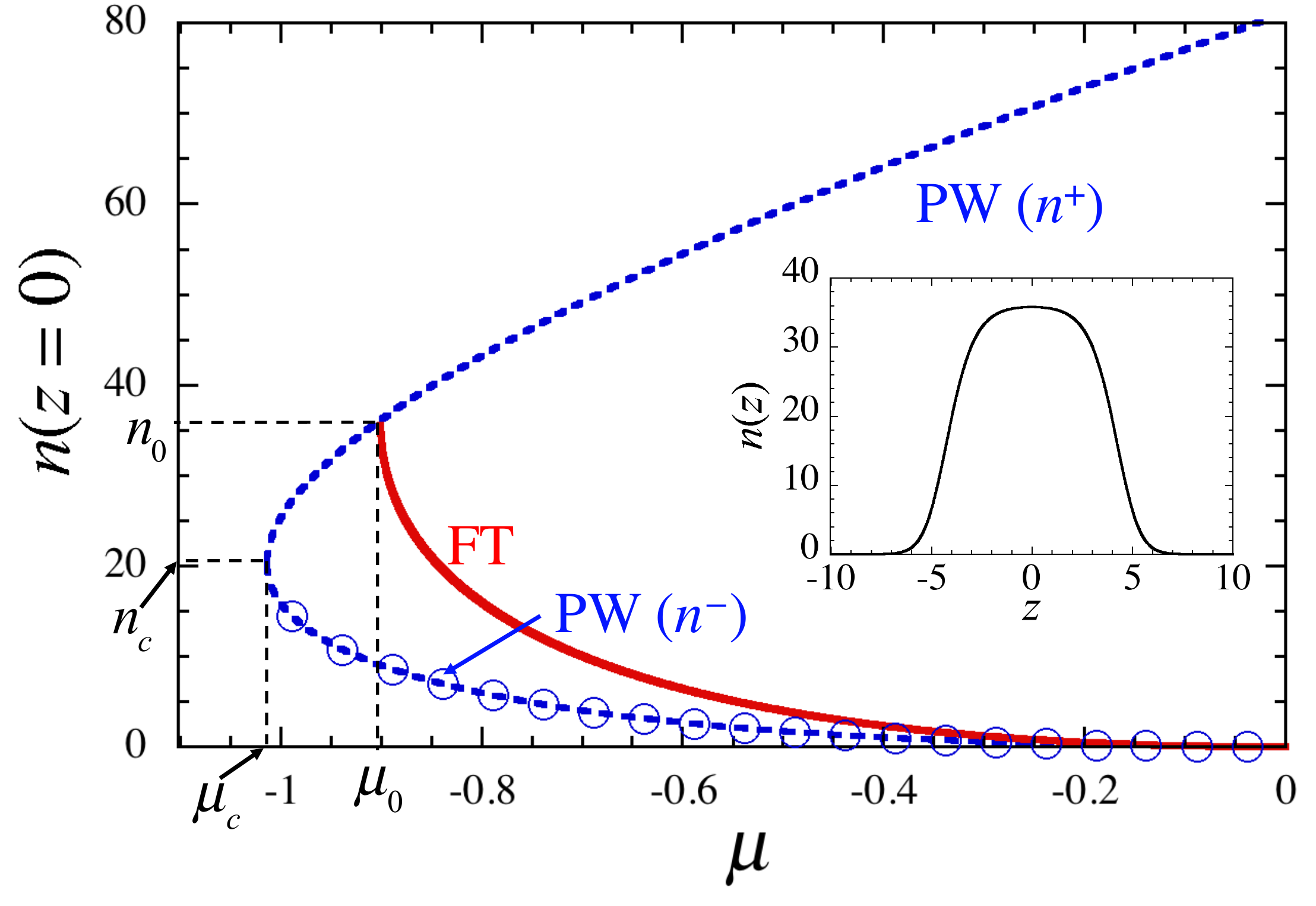}
\caption{The maximum density $n_{\max }\equiv n(z=0)$ in the FT (flat-top)
state, as per Eq.~(\protect\ref{nmax}), and the PW (plane-wave)\ density are
displayed as functions of $\protect\mu $ by\ the red solid and blue dashed
curves, respectively, for $\protect\delta g/g=0.05$. In this case, Eq.~(%
\protect\ref{1n0}) yields $n_{0}=36.025$ and $\protect\mu _{0}=-0.900633$.
The PW solution includes upper and lower branches corresponding to $n^{\pm }$%
, as given by Eq.~(\protect\ref{n}), the lower one (marked by circles) being
subject to the MI (modulational instability). The spinodal point is one with
coordinates $\left( \protect\mu _{c},n_{c}\right) $. For other values of $%
\protect\delta g/g$, the plot can be generated from the present one by
rescaling. The inset shows the density profile of the FT solution for $%
\protect\delta g/g=0.05$ and $\protect\mu =\protect\mu _{0}+0.00001$, very
close to the delocalization limit (the tranistion to PW).}
\label{Kasafig1}
\end{figure}

As shown in Refs.~\cite{Petrov:2016,
Milivoj,Astrakharchik:2018Dynamics}, at $\delta g/g>0$ Eq.~\eqref{eq13}
gives rise to an exact soliton-like solution representing a QD, maintained
by the balance between the effective cubic self-repulsion and quadratic
attraction:

\begin{equation}
\psi (z,t)=\frac{Ae^{-i\mu t}}{1+B\cosh (\sqrt{-2\mu }z)},\quad A=\sqrt{n_{0}%
}\frac{\mu }{\mu _{0}},\quad B=\sqrt{1-\frac{\mu }{\mu _{0}}}.
\label{eq:exact}
\end{equation}%
This solution exists in a finite range of negative values of the chemical
potential $\mu _{0}<\mu <0$, featuring the FT shape at $0<\mu -\mu _{0}\ll
\left\vert \mu _{0}\right\vert $, with size $L\approx \left( -2\mu
_{0}\right) ^{-1/2}\ln \left[ \left( 1-\mu /\mu _{0}\right) ^{-1}\right] $
\cite{Petrov:2016, Astrakharchik:2018Dynamics}. A typical density profile of
the FT solution is displayed in the inset of Fig.~\ref{Kasafig1}. At $\mu
=\mu _{0}$, the size of the droplet diverges, and the solution carries over
into the delocalized PW with uniform density, $n=n_{0}$. The fact that the
density of the condensate filling the FT state cannot exceed the largest
value, $n_{0}$, implies its incompressibility. For this reason, the
condensate may be considered as a fluid, as mentioned above. With the
increase of $\mu $ from $\mu _{0}$ towards $\mu =0$, the maximum density of
the localized mode,

\begin{equation}
n_{\max }\equiv n(z=0)=n_{0}\left( \frac{\mu }{\mu _{0}}\right) ^{2}\left( 1+%
\sqrt{1-\frac{\mu }{\mu _{0}}}\right) ^{-2},  \label{nmax}
\end{equation}%
monotonously decreases from $n_{0}$ to $0$. The QD's FWHM size, defined by
condition $n\left( z=L_{\mathrm{FWHM}}/2\right) =n\left( z=0\right) /2$,
also shrinks at first with increasing $\mu $, attaining a minimum value $%
\left( L_{\mathrm{FWHM}}\right) _{\min }\approx 2.36/\sqrt{-\mu _{0}}$ at $%
\mu /\mu _{0}\approx 0.776$. Further increase of $\mu $ towards $\mu =0$
makes the QD broader, its width diverging as $L_{\mathrm{FWHM}}\approx 1.71/%
\sqrt{-\mu }$ at $\mu \rightarrow -0$.

The norm of the exact QD solutions given by Eqs.~\eqref{eq:exact} is

\begin{equation}
N(\mu )=n_{0}\sqrt{-\frac{2}{\mu _{0}}}\left[ \ln \left( \frac{1+\sqrt{\mu
/\mu _{0}}}{\sqrt{1-\mu /\mu _{0}}}\right) -\sqrt{\frac{\mu }{\mu _{0}}}%
\right] .  \label{NQD}
\end{equation}%
It satisfies the well-known Vakhitov-Kolokolov (VK) necessary stability
criterion \cite{Vakhitov1973stationary},

\begin{equation}
\frac{dN(\mu )}{d\mu }=-\frac{n_{0}}{\mu _{0}^{2}}\sqrt{-\frac{\mu }{2}}%
\frac{1}{1-\mu /\mu _{0}}<0,  \label{Vakh}
\end{equation}%
due to $\mu _{0}<0$ and
\begin{equation}
0<\mu /\mu _{0}<1.  \label{interval}
\end{equation}
Full stability of the QD family has been verified by direct simulations of
the evolution of perturbed QDs in the framework of Eq.~\eqref{eq13}.

It is relevant to mention that exact solution (\ref{eq:exact}) is valid too
at $\delta g/g<0$, when the cubic term in Eq. (\ref{eq13}) is
self-attractive, like the quadratic one. In that case, $\mu _{0}$ is
positive, as per Eq. (\ref{1n0}), while the chemical potential of the
self-trapped state remains negative, as solution (\ref{eq:exact}) may exist
only at $\mu <0$. Then, it follows from Eq. (\ref{eq:exact}) that the
soliton-like mode exists for all values of $\mu <0$ (unlike the finite
interval (\ref{interval}), in which the solution exists for $\delta g/g>0$),
and it does not feature the FT shape. Rather, with the increase of $-\mu $,
it demonstrates a crossover between the KdV-soliton shape $\sim \mathrm{sech}%
^{2}\left( \sqrt{-\mu /2}z\right) $ and the nonlinear-Schr\"{o}dinger one, $%
\sim \mathrm{sech}\left( \sqrt{-2\mu }z\right) $. For $\delta g/g<0$, the $%
N(\mu )$ dependence for the soliton family carries over into the following
form,

\begin{equation}
N(\mu )\biggl|_{\delta g<0}=n_{0}\sqrt{\frac{2}{\mu _{0}}}\left[ \sqrt{-%
\frac{\mu }{\mu _{0}}}-\arctan \left( \sqrt{-\frac{\mu }{\mu _{0}}}\right) %
\right] ,
\end{equation}%
which is an analytical continuation of expression (\ref{NQD}). This
dependence also satisfies the VK criterion.

\subsubsection{The plane-wave solution}

\label{pws} The PW solution of Eq.~(\ref{eq13} )can be presented in a form $%
\psi (z,t)=\sqrt{n}\exp \left( iK_{\text{PW}}z-i\mu t\right) $ with
wavenumber $K_{\text{PW}}$ and constant density $n$, which determine the
corresponding chemical potential:

\begin{equation}
\mu _{\text{PW}}=\frac{\delta g}{g}n-\frac{\sqrt{2}}{\pi }\sqrt{n}+\frac{1}{2%
}K_{\text{PW}}^{2}.  \label{1b}
\end{equation}%
The Galilean invariance of Eq.~(\ref{eq13}) implies that any quiescent
solution $\psi _{0}\left( z,t\right) $ generates a family of moving ones,
with arbitrary velocity $c$. Therefore, $K_{\text{PW}}$ may be canceled by
means of transformation $\psi _{c}\left( z,t\right) =\exp \left(
icz-ic^{2}t/2\right) \psi _{0}\left( z-ct,t\right) $ with $c=-K_{\text{PW}}$.

For given $\mu $, Eq.~(\ref{1b}) produces two different branches of the
density as a function of $\mu $ (here, $K_{\text{PW}}=0$ is set):

\begin{equation}
\sqrt{n^{\pm }(\mu )}=\frac{1}{\sqrt{2}\pi }\frac{g}{\delta g}\pm \sqrt{%
\frac{1}{2\pi ^{2}}\left( \frac{g}{\delta g}\right) ^{2}+\frac{g}{\delta g}%
\mu }.  \label{n}
\end{equation}%
For $\delta g/g=0.05$, these branches are shown in Fig.~\ref{Kasafig1}. As
follows from Eq.~(\ref{n}), they exist (for $\delta g/g>0$) above a minimum
value of $\mu $: $\mu _{c}=-(2\pi ^{2}\delta g/g)^{-1}=(9/8)\mu _{0}$, the
respective density being

\begin{equation}
n_{c}=n^{\pm }(\mu _{c})=\frac{1}{2\pi^2} \left(\frac{g}{\delta g}
\right)^{2} = \frac{9}{16} n_{0}.  \label{nc}
\end{equation}
Values $\mu =\mu _{c}$ and $n=n_{c}$ correspond to the \textit{spinodal
point }\cite{Petrov:2016}, and $n^{+}(\mu _{0})=n_{0}$ (see Eq. (\ref%
{eq:exact})). Note that the\ above-mentioned existence region of the soliton
solution in terms of the chemical potential, $\mu _{0}<\mu <0$, lies
completely inside that of the PW state, which is $\mu _{c}\leq \mu $. Thus,
the soliton always coexists with the PW (this fact is also obvious in Fig. %
\ref{Kasafig1}).

\subsubsection{Modulational instability of the plane waves}

\label{MIsingled} Here, we aim to analyze the MI of PW solutions in the
framework of the single-component GP equation (\ref{eq13}) and demonstrate
how the development of the MI can help to generate QDs. We perform the
analysis for the PWs with zero wavenumber $K_{\text{PW}}=0$, which is
sufficient due to the aforementioned Galilean invariance of the underlying
equation.

A small perturbation is added to the stationary PW state as

\begin{equation}
\psi (z,t)=\left[ \sqrt{n}+\delta \psi (z,t)\right] \exp \left( -i\mu
t\right) .  \label{sing-pert}
\end{equation}%
The substitution of this expression in Eq.~(\ref{eq13}) and linearization
with respect to perturbation $\delta \psi $ leads to the corresponding
Bogoliubov-de Gennes equation,

\begin{equation}
i\frac{\partial }{\partial t}\delta \psi =-\frac{1}{2}\frac{\partial ^{2}}{%
\partial z^{2}}\delta \psi +\frac{\delta g}{g}n(\delta \psi +\delta \psi
^{\ast })-\frac{\sqrt{n}}{\sqrt{2}\pi }(\delta \psi +\delta \psi ^{\ast }).
\label{eigen}
\end{equation}%
By looking for perturbation eigenmodes with wavenumber $k$ and frequency $%
\Omega $,

\begin{equation}
\delta \psi =\zeta \cos (kz-\Omega t)+i\eta \sin (kz-\Omega t),
\label{pert-form}
\end{equation}%
and real infinitesimal amplitudes $\zeta $ and $\eta $, Eq.~(\ref{eigen})
yields a dispersion relation for the eigenfrequencies:

\begin{equation}
\Omega ^{2}=\frac{k^{4}}{4}+\left( \frac{\delta g}{g}n-\frac{\sqrt{n}}{\sqrt{%
2}\pi }\right) k^{2}.  \label{pert-eig}
\end{equation}

\begin{figure}[tbh]
\centering
\includegraphics[width=0.95\columnwidth]{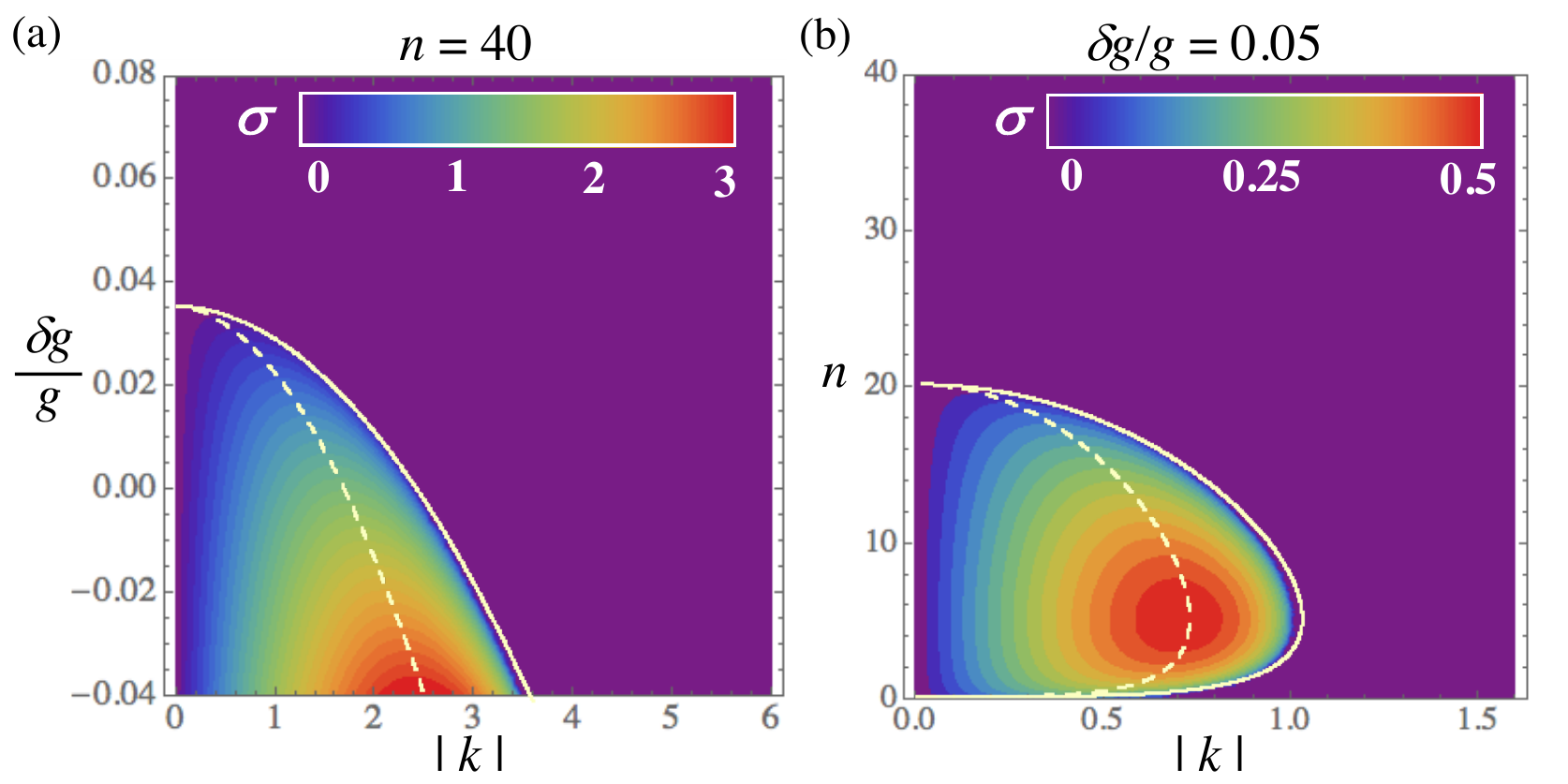}
\caption{Color-coded values of the MI gain, $\protect\sigma =\mathrm{Im}%
(\Omega )$, are displayed for fixed $n=40$ in (a), and for fixed $\protect%
\delta g/g=0.05$ in (b). Note that panel (a) covers both signs of the cubic
nonlinearity, $\protect\delta g>0$ and $\protect\delta g<0$. Solid and
dashed white curves represent the MI boundary [Eq.~\eqref{k0}] and the peak
value of the MI gain [Eq.~\eqref{kmax}], respectively.}
\label{Fig. rel_freq2}
\end{figure}
The MI takes place when $\Omega $ acquires an imaginary part. As follows
from Eq.~\eqref{pert-eig}, this occurs when the density satisfies condition $%
n<[2\pi ^{2}(\delta g/g)^{2}]^{-1}=n_{c}$ [see Eq. (\ref{nc})], which
corresponds to branch $n^{-}$ of the PW state. The instability region in
terms of $k$ is given by

\begin{equation}
k^{2}<4\left( \frac{\sqrt{n}}{\sqrt{2}\pi }-\frac{\delta g}{g}n\right)
\equiv k_{0}^{2}.  \label{k0}
\end{equation}%
The MI gain $\sigma \equiv \left\vert \mathrm{Im}\Omega \right\vert $ is
plotted in Fig.~\ref{Fig. rel_freq2} versus $|k|$ and $\delta g/g$, for
given density $n=40$ in panel (a), and versus $|k|$ and $n$, for given $%
\delta g/g=0.05$ in (b). It is easy to find from Eq.~(\ref{pert-eig}) that
the largest gain is attained at wavenumber

\begin{equation}
k_{\max }=\frac{k_{0}}{\sqrt{2}},  \label{kmax}
\end{equation}%
with $k_{0}$ defined as per Eq.~(\ref{k0}). Note that Fig.~\ref{Fig.
rel_freq2}(a) includes the case of the self-attractive cubic nonlinearity,
with $\delta g/g<0$, which naturally displays much stronger MI, as in this
case it is driven by both the quadratic and cubic nonlinear terms. In fact,
the extension of the MI chart to $\delta g/g<0$ makes it possible to
complare the MI in the present system and its well-known counterpart in the
setting with the fully attractive nonlinearity.

\begin{figure}[tbh]
\centering
\includegraphics[width=0.95\columnwidth]{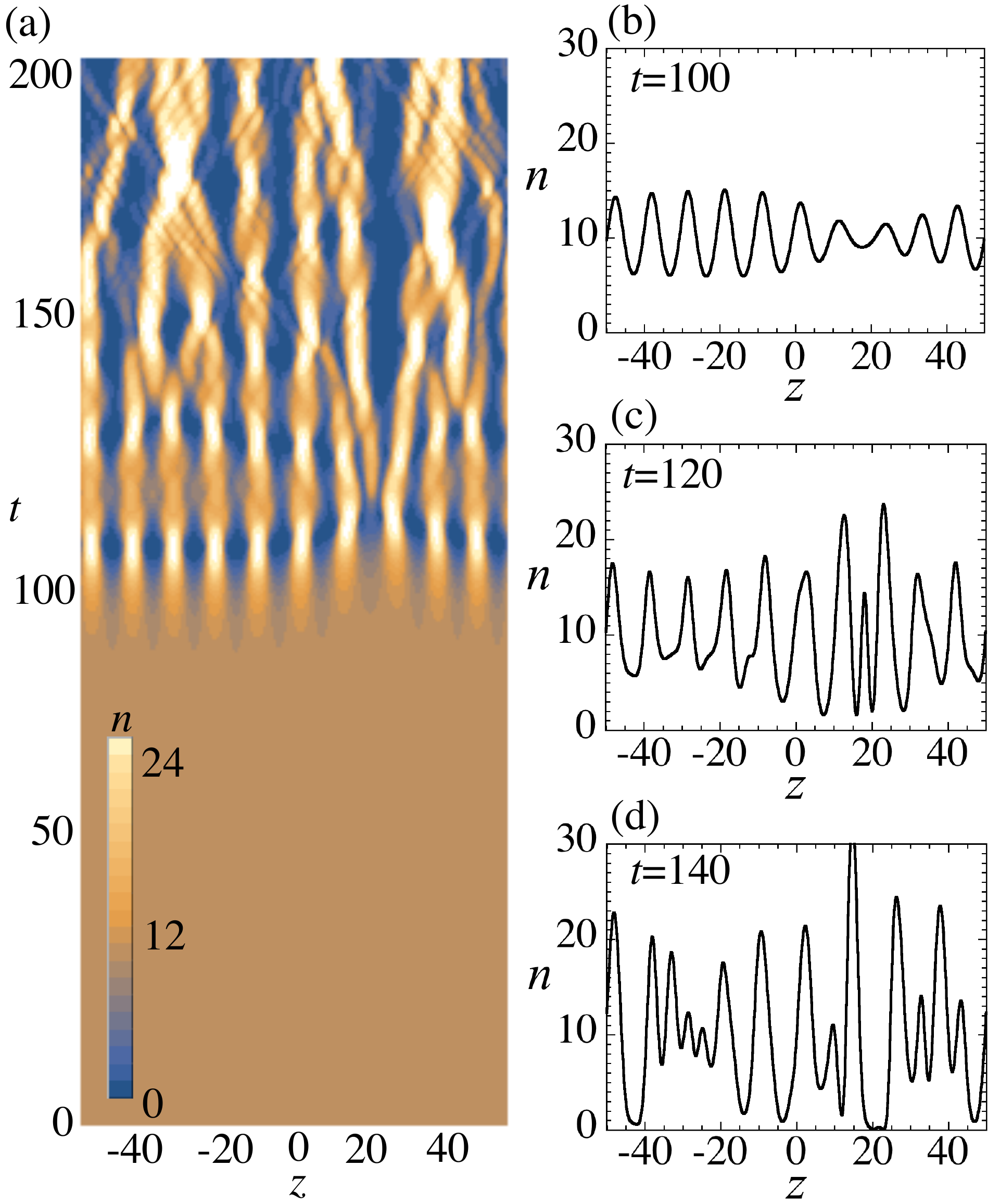}
\caption{A typical example of the MI development, starting from an unstable
PW state, with density $n=10$ and $\protect\delta g/g=0.05$, which is
subject to the MI, pursuant to Fig.~\protect\ref{Fig. rel_freq2}. In (a),
the spatiotemporal pattern of the evolution of the condensate density is
shown. In the right-hand panels, cross sections of the density profiles are
displayed at $t=100$ (b), $t=120$ (c), and $t=140$ (c). The simulations were
performed in domain $-50<z<+50$ with $2500$ grid points and periodic
boundary conditions.}
\label{homodyn}
\end{figure}

Comparing parameter values at which the QD solutions are predicted to
appear, and the MI condition for the PW with the corresponding density, the
MI is expected to provide a mechanism for the creation of the QDs. This is
confirmed by direct simulations of the GP equation \eqref{eq13}, as shown in
Fig.~\ref{homodyn}. The PW with $n=10$ is taken as the input, so that it is
subject to the MI for $\delta g/g=0.05$, as seen in Fig.~\ref{Fig. rel_freq2}%
(b). As shown in Fig.~\ref{homodyn}, small initial perturbations trigger the
emergence of multiple-QD patterns (chains) at $t\geq 100$. For these
parameters, we get $k_{\text{max}}\ =0.6508$ and $\sigma \left( k_{\text{max}%
}\right) =0.2118$, which determines the wavelength of the fastest growing
modulation, $\lambda =2\pi /k_{\text{max}}\ \approx 9.66$, and the
growth-time scale, $\tau =2\pi /\sigma \left( k_{\text{max}}\right) \approx
30$. The number of the generated droplets in Fig.~\ref{homodyn} is
consistent with estimate $L/\lambda \simeq 10$, where $L=100$ is the size of
the simulation domain. We have checked that the number of generated droplets
is approximately given by $L/\lambda $ for other values of parameters as
well. This dynamical scenario is similar to those observed in other models
in the course of the formation of soliton chains by MI of PWs \cite%
{Nguyen:2017Formation,Everitt:2017Observation}. The long-time evolution in
Fig.~\ref{homodyn}(a) shows that the number of the droplets becomes smaller
due to merger of colliding droplets into a single one, which agrees with
dynamical properties of 1D QDs reported in Ref.~\cite%
{Astrakharchik:2018Dynamics}.

To implement this mechanism of the generation of a chain of solitons in the
experiment, i.e., make the density smaller than the critical value $n_c$,
one may either apply interaction quench (by means of the Feshbach
resonance), suddenly decreasing $\delta g/g$, as was done in recent
experimental works for different purposes \cite%
{cabrera2018quantum,Cheiney:2018Bright,Semeghini:2018Self,Strathclyde}.
Another option, which is specific to the 1D setting, is sudden decrease of
density $n$ by relaxing the transverse trapping.


\subsection{The two-component Gross-Pitaevskii model}

{\label{sec3}} In this section, we revert to the full two-component GP
system \eqref{2comp_GP}, aiming to explore the formation of QD states in it.
The two-component setting may include parameter imbalance between the two
components, as indicated theoretically \cite{Petrov:2015} and observed
experimentally \cite%
{cabrera2018quantum,Cheiney:2018Bright,Semeghini:2018Self,long-lived}. Here,
we present the analysis of asymmetric QDs in two cases: (i) the
two-component GP system with different populations, $N_{1}/N_{2}\neq 1$, and
equal intra-component coupling strength, $g_{1}=g_{2}$ (i.e., $P=1$, see Eq.
(\ref{PG})), and (ii) the system with different intra-component coupling
strengths, $g_{1}\neq g_{2}$ (i.e., $P\neq 1$). These options suggest a
possibility to check the stability of the solutions of the symmetric system,
reduced to the single-component form, against symmetry-breaking
perturbations. That objective is relevant because, in the real experiment,
scattering lengths of the self-interaction in the two components are never
exactly equal \cite{cabrera2018quantum}-\cite{2018arXiv181209151F}. We
address, first, an asymmetric single-droplet solution, and, subsequently, MI
of the PW states in the two-component system.

Because, as said above, solutions with mutually proportional components
(written as $\phi _{1}=K\phi _{2}$) are possible solely in the strictly
symmetric setting, asymmetric QDs cannot be found in an exact analytical
form. As shown in Appendix B [see Eqs.~(\ref{phi1})-(\ref{dip})], asymptotic
analytical solutions can be obtained for strongly asymmetric states, with
one equation replaced by its linearized version. In this section, we chiefly
rely on numerical solution of Eq.~\eqref{2comp_GP}.

\subsubsection{Asymmetric QDs with unequal populations ($N_{1}\neq N_{2}$)
for $g_1 = g_2$ ($P=1$)}

\label{sympopudif} In the system with $P=1$ [see Eq. (\ref{PG})], we
calculated the droplet states as stationary solutions of Eq.~\eqref{2comp_GP}
by means of the imaginary-time-evolution method with the Neumann's boundary
conditions, under the constraint that the norm is fixed in the first
component, $\int_{-\infty }^{+\infty }dz|\psi _{1}(z)|^{2}=N_{1}$, while
chemical potential $\mu _{2}$ is fixed in the other one, allowing its norm $%
N_{2}$ to vary.

\begin{figure}[tbh]
\centering
\includegraphics[width=1.0\columnwidth]{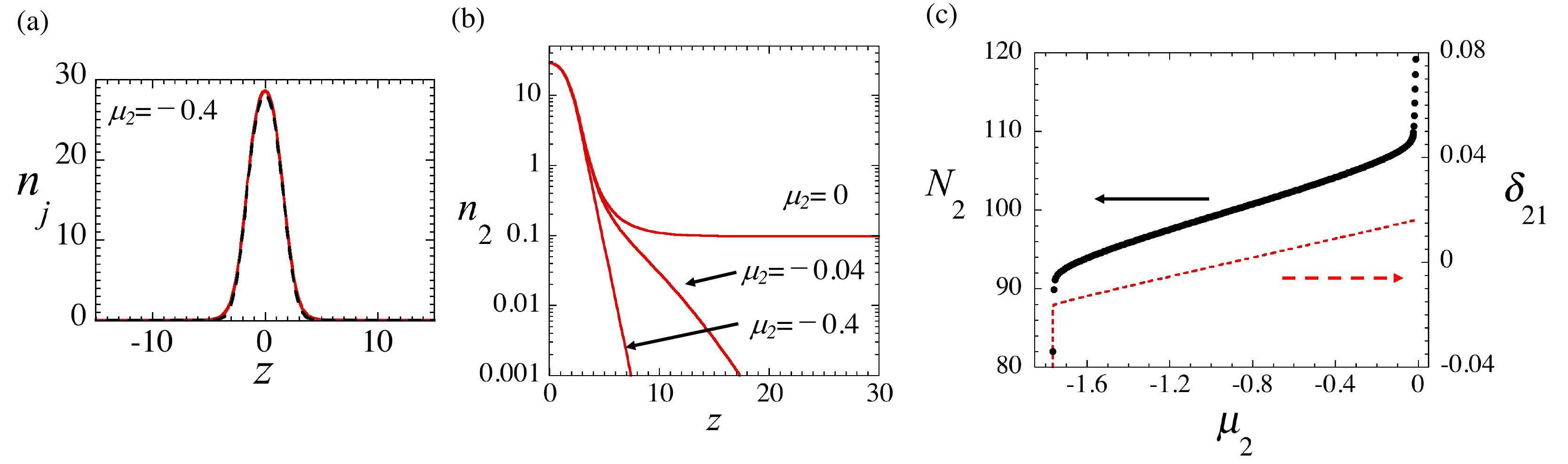}
\caption{(a) Stationary weakly asymmetric (with respect to the two
components) solutions of Eq.~\eqref{2comp_GP}, obtained for $\protect\mu %
_{2}=-0.4$ with fixed $N_{1}=100$. Dashed and solid curves display density
profiles of the first ($n_{1}$) and second ($n_{2}$) components,
respectively. (b) The semi-log plot of the density profiles of $n_{2}$ for $%
\protect\mu _{2}=-0.4$, $-0.04$, and $0$ at $z>0$. (c) Dependences of $N_{2}$
(black dots: the left vertical axis) and asymmetry parameter $\protect\delta %
_{21}$, defined as per Eq. (\protect\ref{delta12asy}) (the red dashed line
pertaining to the right vertical axis), on $\protect\mu _{2}$ for fixed $%
N_{1}=100$. The parameters are $P=1$ $(g_{1}=g_{2})$ and $\protect\delta %
g/g=0.05$. The symmetric point with $N_{1}=N_{2}=100$ and $\protect\delta %
_{21}=0$ corresponds to $\protect\mu _{1}=\protect\mu _{2}=-0.88878$. }
\label{Fig_2ground_as1}
\end{figure}

Figure~\ref{Fig_2ground_as1} displays essential features of weakly
asymmetric droplets for $\delta g/g=0.05$ and fixed $N_{1}=100$. The
symmetric (completely overlapping) solution with $N_{1}=N_{2}$ is found at $%
\mu _{1}=\mu _{2}=-0.88878$. When $\mu _{2}$ deviates from this value,
profiles of the two components become slightly different, as shown in Fig.~%
\ref{Fig_2ground_as1}(a). The profiles of the droplet solution hardly change
for different values of $\mu_2$, but panel \ref{Fig_2ground_as1}(b)
demonstrates that, at $\mu_2 \to -0$, $\psi _{2}$ develops small-amplitude
extended tails, which are absent in $\psi _{1}$. Due to the contribution of
the tails, the approach of $\mu _{2}<0$ towards zero leads to the increase
of norm $N_{2}$, as seen in Fig.~\ref{Fig_2ground_as1}(c). Note that the
growth of $N_{2}(\mu _{2})$ at $\mu _{2}\rightarrow -0$ is opposite to the
decay of the QD's norm in the single-component model at $\mu \rightarrow -0$%
, cf. Eq.~(\ref{NQD}). At $\mu _{2}\geq 0$, the $\psi _{2}$ component
undergoes delocalization, with its tails developing a nonzero background at $%
|z|\rightarrow \infty $, as seen in the density profile displayed in Fig.~%
\ref{Fig_2ground_as1}(b) for $\mu _{2}=0$, and norm $N_{2}(\mu_2 )$
diverging at $\mu_2 \rightarrow -0$ in Fig.~\ref{Fig_2ground_as1}(c).

In Fig.~\ref{Fig_2ground_as1}(c), we also plot the parameter of the
asymmetry between the two components, defined as

\begin{equation}
\delta _{21}=\frac{n_{2}(z=0)-n_{1}(z=0)}{n_{2}(z=0)+n_{1}(z=0)}.
\label{delta12asy}
\end{equation}%
It increases almost linearly with $\mu _{2}$, although its absolute value
does not exceed $0.02$. Thus, the droplet tends to keep a nearly symmetric
profile, with respect to the two components, in the symmetric system, even
if the population imbalance is admitted. In fact, this circumstance makes
the analysis self-consistent, as the use of the GP system with the LHY
correction implies that the MF intra- and inter-component interactions
nearly cancel each other, which is possible only if shapes of the two
components are nearly identical.

\subsubsection{Asymmetric QDs in the system with $P\neq 1$ ($g_{1}\neq g_{2}$%
)}

\label{qdasssym} Next, we consider the QDs for $P\neq 1$, setting $P>1$
without loss of generality. Then, the MF energy is minimized for $%
n_{2}>n_{1} $; the situation with $n_{1}>n_{2}$ can be considered too,
replacing $P$ by $P^{-1}$.

\begin{figure}[tbh]
\centering
\includegraphics[width=0.95\columnwidth]{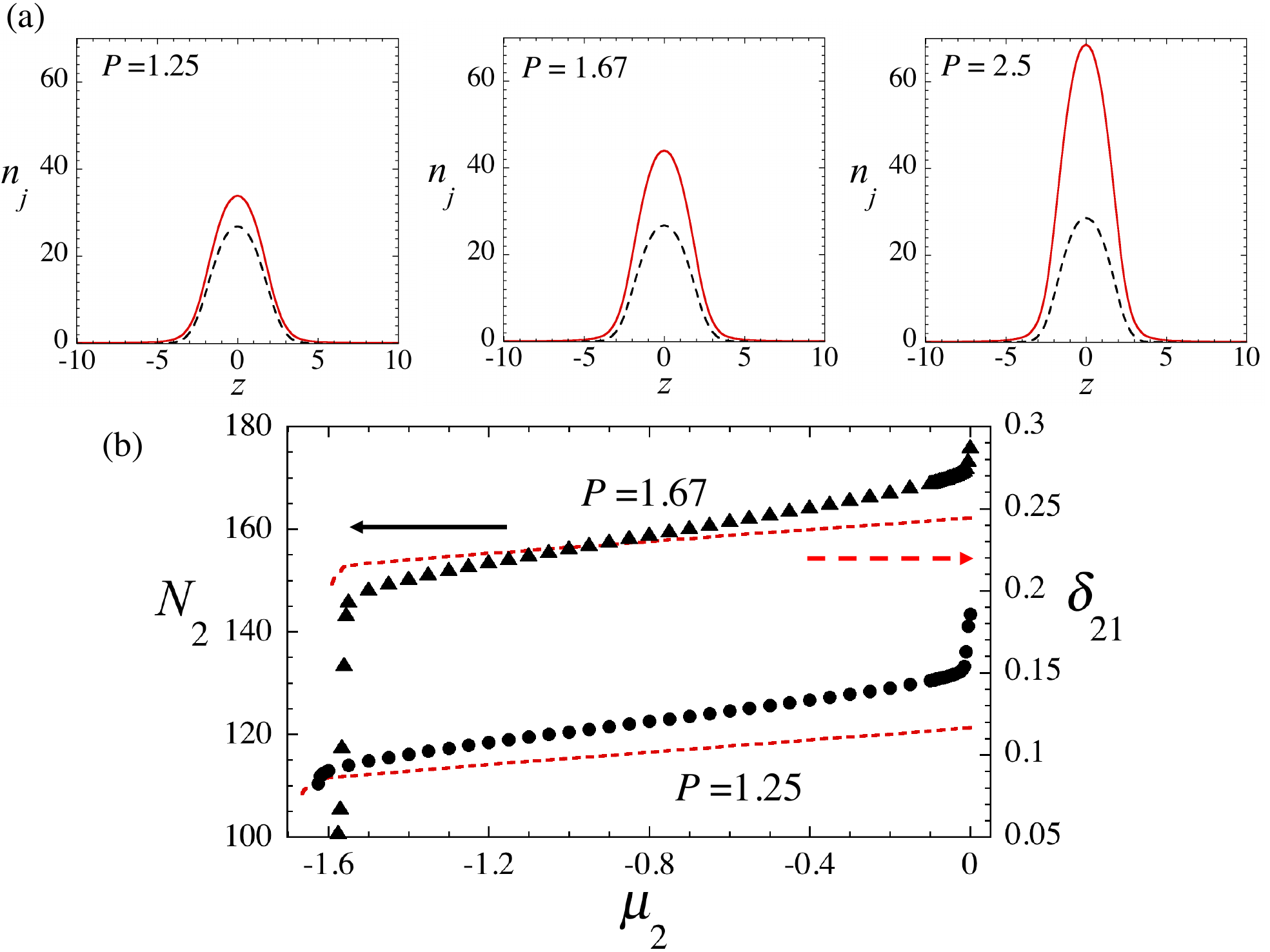}
\caption{(a) Stationary solutions of Eq.~\eqref{2comp_GP}, obtained for $%
\protect\delta g/g=0.05$ and $N_{1}=100$. From the left panel to the right
one, the parameter (\protect\ref{PG}) is $P=1.25$, $1.67$, and $2.5$, and
the chemical potential for the second component is $\protect\mu _{2}=-0.018$%
, $-0.011$, and $-0.006$, respectively, just below the threshold above which
the tails of $\protect\psi _{2}$ extend to infinity. Dashed and solid curves
represent the density of the first ($n_{1}$) and second ($n_{2}$)
components. (b) Dependences of $N_{2}$ (black dots: the left vertical axis)
and asymmetry parameter $\protect\delta _{21}$, defined as per Eq. (\protect
\ref{delta12asy}) (the red dashed line pertaining to the right vertical
axis), on $\protect\mu _{2}$ for fixed $N_{1}=100$ and $P=1.25$ or $P=1.67$.
}
\label{Fig_difcritical2}
\end{figure}

Following the procedure similar to that employed in Sec.~\ref{sympopudif},
we produce QD solutions for $\delta g/g=0.05$, $N_{1}=100$, and several
different values of $P$, varying $\mu _{2}$. In Fig.~\ref{Fig_difcritical2}
(a), we plot density profiles for three different values of $P$. Naturally,
the difference of the two components increases with the increase of $P$. In
Fig.~\ref{Fig_difcritical2}(b) we display $N_{2}$ and parameter $\delta
_{21} $ [see Eq. (\ref{delta12asy})] of the asymmetric QDs for $P=1.25$ and $%
1.67$. All these states have been checked to be stable in time-dependent
simulations.

\begin{figure}[tbh]
\centering
\includegraphics[width=0.95\columnwidth]{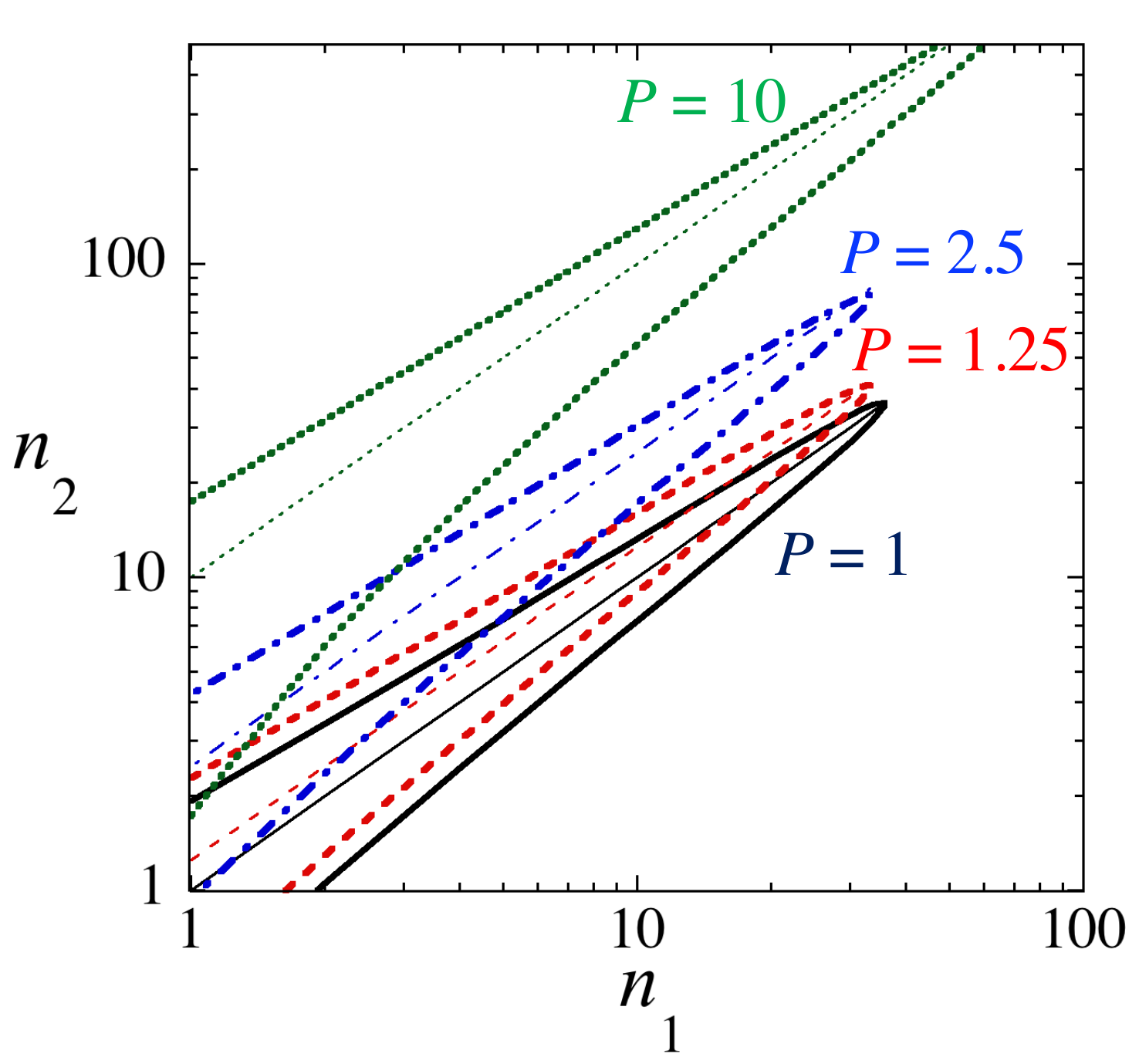}
\caption{The negative-pressure region in the $\left( n_{1},n_{2}\right) $
plane for $\protect\delta g/g=0.05$ and values of asymmetry parameter (
\protect\ref{PG}) $P=1$ (the solid curve), $1.25$ (dashed), $2.5$
(dashed-dotted), and $10$ (dotted). Boundaries are determined by the
zero-pressure condition, as given by Eq.~\eqref{zeropressurecond}. The
negative pressure, at which localized states may exist, occurs inside the
boundaries. Thin lines represent relation $n_{2}=Pn_{1}$.}
\label{Figlda}
\end{figure}

The density difference at the center of the droplet can be determined by the
condition of the existence of the liquid phase in the free space. This
condition is obtained by minimizing the grand-potential density $\mathcal{E}%
_{\text{1D}}\ -\mu _{1}\rho _{1}-\mu _{2}\rho _{2}$ \cite%
{Petrov:2016,Ancilotto2018}, which leads to the zero-pressure condition,

\begin{align}
p(\rho _{1},\rho _{2})& =-\mathcal{E}_{\mathrm{1D}}+\sum_{j=1,2}\left( \frac{%
\partial \mathcal{E}_{\mathrm{1D}}}{\partial \rho _{j}}\right) \rho _{j}
\notag \\
& \equiv -\mathcal{E}_{\text{\textrm{1D}}}\ +\mu _{1}\rho _{1}+\mu _{2}\rho
_{2}=0.
\end{align}%
From this, we obtain relation

\begin{equation}
\frac{(\sqrt{g_{1}}\rho _{1}-\sqrt{g_{2}}\rho _{2})^{2}}{2}+\frac{g\delta g(%
\sqrt{g_{2}}\rho _{1}+\sqrt{g_{1}}\rho _{2})^{2}}{(g_{1}+g_{2})^{2}}-\frac{%
\sqrt{m}}{3\pi \hbar }(g_{1}\rho _{1}+g_{2}\rho _{2})^{3/2}=0,
\label{2dropletequicond}
\end{equation}%
which can be rewritten in the scaled form as

\begin{equation}
\frac{P+GP^{-1}}{2}n_{1}^{2}+\frac{P^{-1}+GP}{2}n_{2}^{2}+(G-1)n_{1}n_{2}=%
\frac{1}{3\pi }\left( Pn_{1}+\frac{n_{2}}{P}\right) ^{3/2}.
\label{zeropressurecond}
\end{equation}%
For given $n_{1}$, we solved Eq.~(\ref{zeropressurecond}) to find the
respective value of $n_{2}$, which is shown in Fig.~\ref{Figlda} for $\delta
g/g=0.05$ and several values of $P$. There are two branches of the
solutions, that enclose the negative-pressure region, in which QDs may
exist. The maximum value of $n_{j}$ at the tip of the negative-pressure
region corresponds to the density in the droplet's FT segment. The ascending
negative-pressure region for each $P$ nearly follows relation $n_{2}=Pn_{1}$%
, which is derived by the minimization condition for the dominant first term
in Eq.~\eqref{2dropletequicond} It is seen that a larger difference in the
profiles of the two components occurs for larger $P$, as expected. Also, for
given $n_{1}$, the negative-pressure region becomes wider with respect to $%
n_{2}$ for larger $P$ (note that the figure displays a log-log plot).

\begin{figure}[tbh]
\centering
\includegraphics[width=1.0\columnwidth]{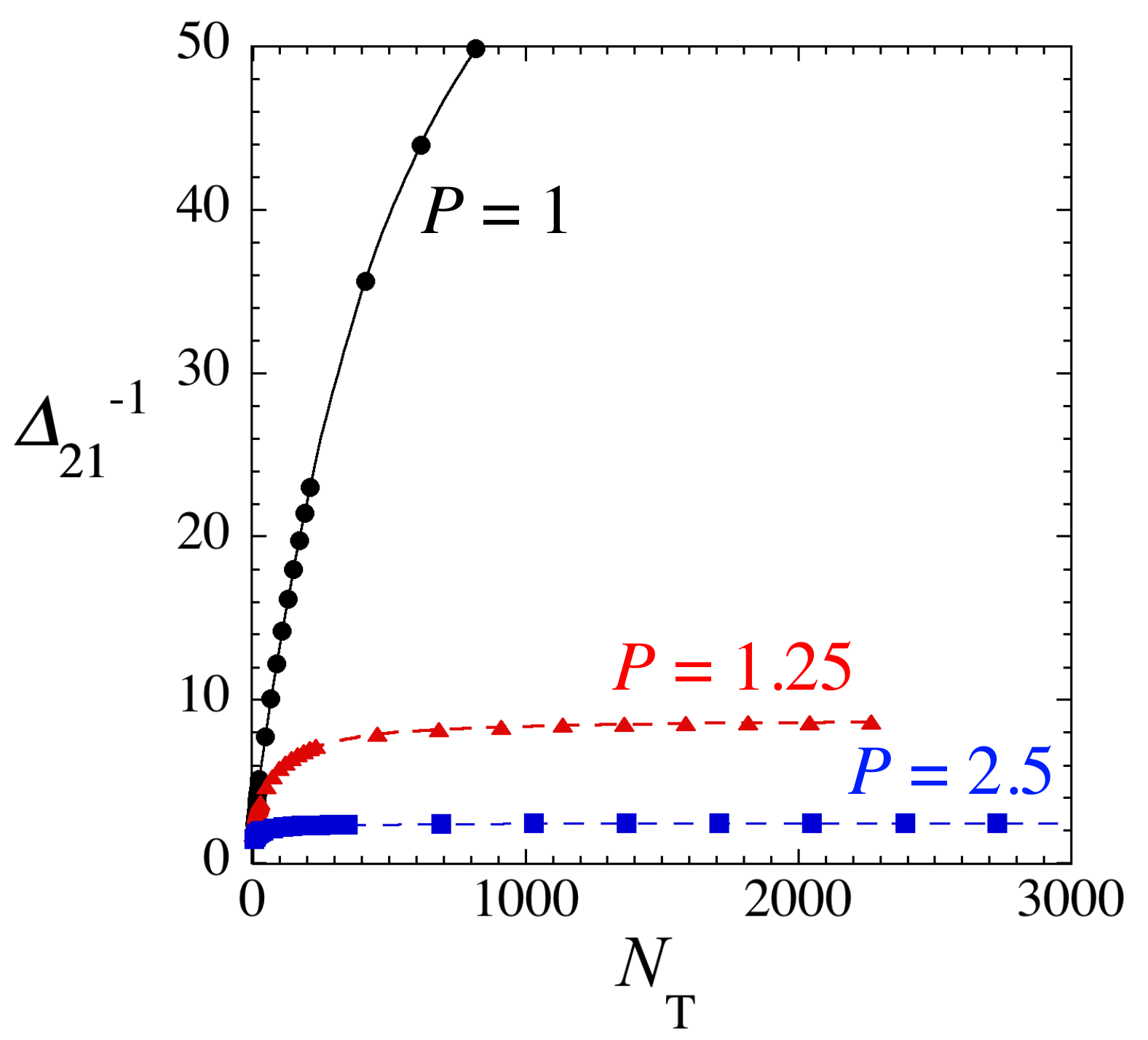}
\caption{The inverse of the largest relative norm difference $\Delta _{21}$,
up to which the asymmetric droplets exist [see Eq. (\protect\ref{Delta})],
shown as a function of the total number, $N_{\mathrm{T}}$, at different
values of asymmetry parameter (\protect\ref{PG}). Here we set $\protect%
\delta g/g=0.05$. }
\label{Fig_difcritical}
\end{figure}
As the QDs have a finite norm, it is relevant to characterize the asymmetry
in terms of the norm, rather than density. Here, we aim to find a largest
value of the norm difference,

\begin{equation}
\Delta _{21}=(N_{2}-N_{1})/N_{\mathrm{T}},  \label{Delta}
\end{equation}%
where $N_{\mathrm{T}}=N_{1}+N_{2}$ is the total norm, which admits the
existence of the QDs. For given $N_{1}$, we obtain the upper bound for $%
N_{2} $ above which the solution becomes delocalized, and calculate the
corresponding critical value of $\Delta _{21}$. The results are shown in
Fig.~\ref{Fig_difcritical}. For the system with $P=1$ , the curve
demonstrates an empirical dependence $\Delta _{21}\propto N_{\mathrm{T}%
}^{-\alpha }$ with exponent $\alpha \approx 0.58$. Accordingly, the
asymmetry tends to vanish asymptotically for very \textquotedblleft heavy"
droplets, at $N_{\mathrm{T}}~\rightarrow \infty $. As the system becomes
slightly asymmetric, with $P=1.25$, exponent $\alpha $ is significantly
reduced for small $N_{\text{\textrm{T}}}$, and converges to a certain finite
value at $N_{\text{\textrm{T}}}\ \rightarrow \infty $. Thus, it is again
confirmed that values $P>1$ maintain conspicuous asymmetry between the QD's
components. Finally, strongly asymmetric non-FT (Gaussian-shaped \cite%
{Astrakharchik:2018Dynamics}) solutions can be obtained in an approximate
analytical form for any value of $P$, as shown in Appendix B.

\subsubsection{The MI of the asymmetric PW states}

The MI of two-component asymmetric PWs is a relevant subject too. Such
solutions are written as $\psi _{j}(z,t)=\sqrt{n_{j}}e^{-i\mu
_{j}t},\,(j=1,2)$. The substitution of this in Eq.~(\ref{2comp_GP}) yields

\begin{eqnarray}
\mu _{1} &=&(P+GP^{-1})n_{1}+(-1+G)n_{2}-\frac{P}{\pi }\sqrt{Pn_{1}+\frac{%
n_{2}}{P}},~~~  \notag \\
\mu _{2} &=&(P^{-1}+GP)n_{2}+(-1+G)n_{1}-\frac{1}{\pi P}\sqrt{Pn_{1}+\frac{%
n_{2}}{P}}.  \label{w-pert2}
\end{eqnarray}%
Accordingly, in the symmetric system with $P=1$, densities of the asymmetric
PW state are expressed in terms of the chemical potentials as

\begin{equation}
\begin{split}
n_{j}&=\frac{1}{4}\left[ \frac{1}{\pi ^{2}G^{2}}+\frac{\mu _{1}+\mu _{2}}{G}%
+(-1)^{j+1}(\mu _{1}-\mu _{2})\right] \\& \pm \frac{\sqrt{1+2\pi ^{2}G(\mu
_{1}+\mu _{2})}}{4\pi ^{2}G^{2}}.
\end{split}
\end{equation}%
We introduce the perturbation around the PW states as

\begin{equation}
\psi _{j}(z,t)=\left[ \sqrt{n_{j}}+\delta \psi _{j}(z,t)\right] e^{-i\mu
_{j}t},  \label{sing-pert2}
\end{equation}%
\begin{equation}
\delta \psi _{j}=\zeta _{j}\cos (kz-\Omega t)+i\eta _{j}\sin (kz-\Omega t),
\label{pert-form3}
\end{equation}%
with infinitesimal amplitudes $\zeta _{j}$ and $\eta _{j}$, cf. Eq. (\ref%
{pert-form}). The substitution of this in Eqs.~(\ref{2comp_GP})\ and the
linearization with respect to $\zeta _{j}$ and $\eta _{j}$ yields the
dispersion equation for the perturbation:

\begin{equation}
\begin{split}
\Omega _{\pm }^{2}=\frac{k^{2}}{4}\left[ k^{2}+2(P_{1}+P_{2}-Q_{1}-Q_{2})%
\right] \\ \pm \frac{k^{2}}{2}\sqrt{(P_{1}-P_{2}-Q_{1}+Q_{2})^{2}+4(R-S)^{2}},
\label{MI2gain}
\end{split}
\end{equation}%
where

\begin{gather}
P_{1}=(P+GP^{-1})n_{1},\quad P_{2}=(P^{-1}+GP)n_{2},  \notag \\
Q_{1}=\frac{P^{2}n_{1}}{2\pi \sqrt{Pn_{1}+P^{-1}n_{2}}},\quad Q_{2}=\frac{%
P^{-2}n_{2}}{2\pi \sqrt{Pn_{1}+P^{-1}n_{2}}}. \\
R=(-1+G)\sqrt{n_{1}n_{2}},\quad S=\frac{\sqrt{n_{1}n_{2}}}{2\pi \sqrt{%
Pn_{1}+P^{-1}n_{2}}},  \notag
\end{gather}%
For $P=1$ and $n_{1}=n_{2}$, these results reproduce Eq.~\eqref{pert-eig}
for the $\Omega _{-}$ branch. A parameter region in which at least one
squared eigenfrequency $\Omega _{\pm }^{2}$ is negative gives rise to the MI
of the two-component state.

\subsubsection{The MI for $P=1$}

\begin{figure*}[tbh]
\centering
\includegraphics[width=1.0\linewidth]{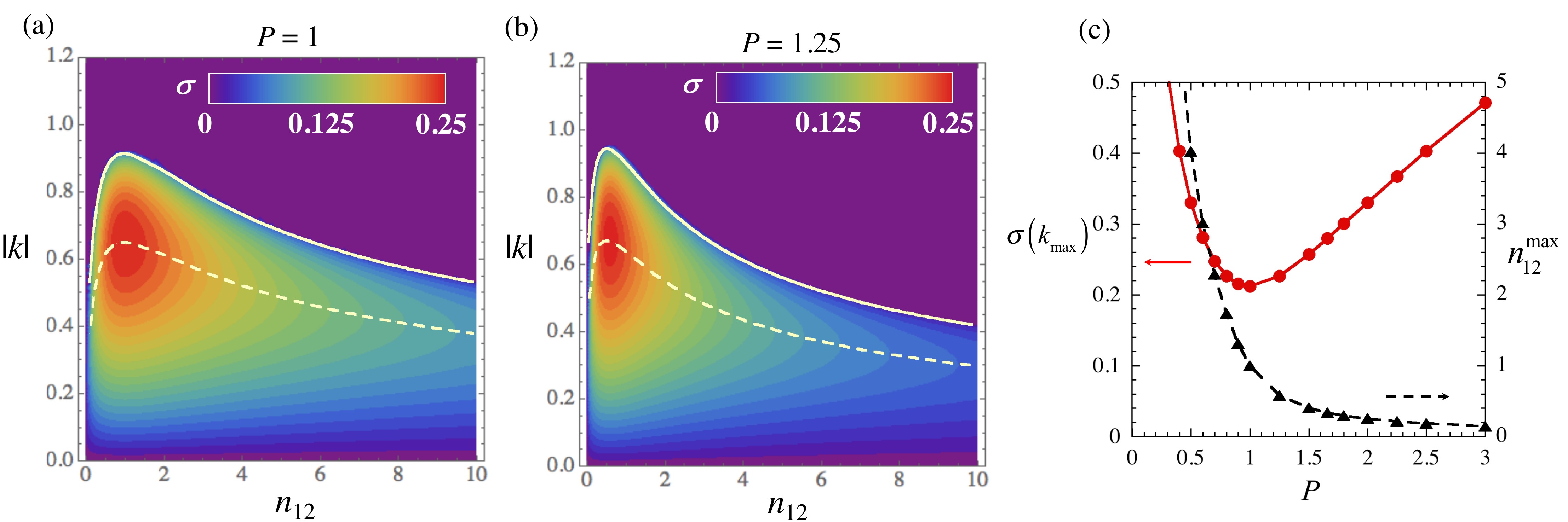}
\caption{Color-coded values of the MI gain, $\protect\sigma =\mathrm{Im}%
(\Omega )$, for asymmetric PWs, as calculated from Eq.~\eqref{MI2gain} in
the plane of wave number $|k|$ and density ratio $n_{12}=n_{1}/n_{2}$, are
displayed for (a) $P=1$ and (b) $P=1.25$ with fixed $\protect\delta g/g=0.05$
and $(n_{1}+n_{2})/2=10$. The solid and dashed white curves represent the MI
boundary $k=k_{0}$ and the peak value of the MI gain at $k=k_{\text{max}}\
=k_{0}/\protect\sqrt{2}$, respectively. In (c), we plot $\protect\sigma (k_{%
\text{max}})$ (circles) and $n_{12}^{\text{max}}$ (triangles) versus $P$.}
\label{FigMItwo}
\end{figure*}

In Fig.~\ref{FigMItwo}, we plot the gain spectrum $\sigma =\mathrm{Im}%
(\Omega )$ for the asymmetric PWs in the symmetric system with $P=1$ and $%
\delta g/g=0.05$, in the plane of wavenumber $k$ and density ratio $%
n_{12}=n_{2}/n_{2}$. For the consistency with the single-component situation
displayed in Fig.~\ref{homodyn}, we here fix the total density as $%
(n_{1}+n_{2})/2=10$. For given $n_{12}$, the MI occurs at $|k|<k_{0}$, and
the gain attains its maximum at $k=k_{\text{max}}=k_{0}/\sqrt{2}$. The
largest gain is obtained at equal densities, $n_{12}=1$. Both the $k$-band
of the instability and magnitude of the gain slowly decrease as the
deviation of $n_{12}$ from unity increases. This means that the MI occurs in
the PW states with a large density difference, thus giving rise to the
formation of solitons with large asymmetry even for equal intra-component MF
interaction strengths, $P=1$ [see Eq. (\ref{PG})] .

\begin{figure}[!tbh]
\centering
\includegraphics[width=0.95\columnwidth]{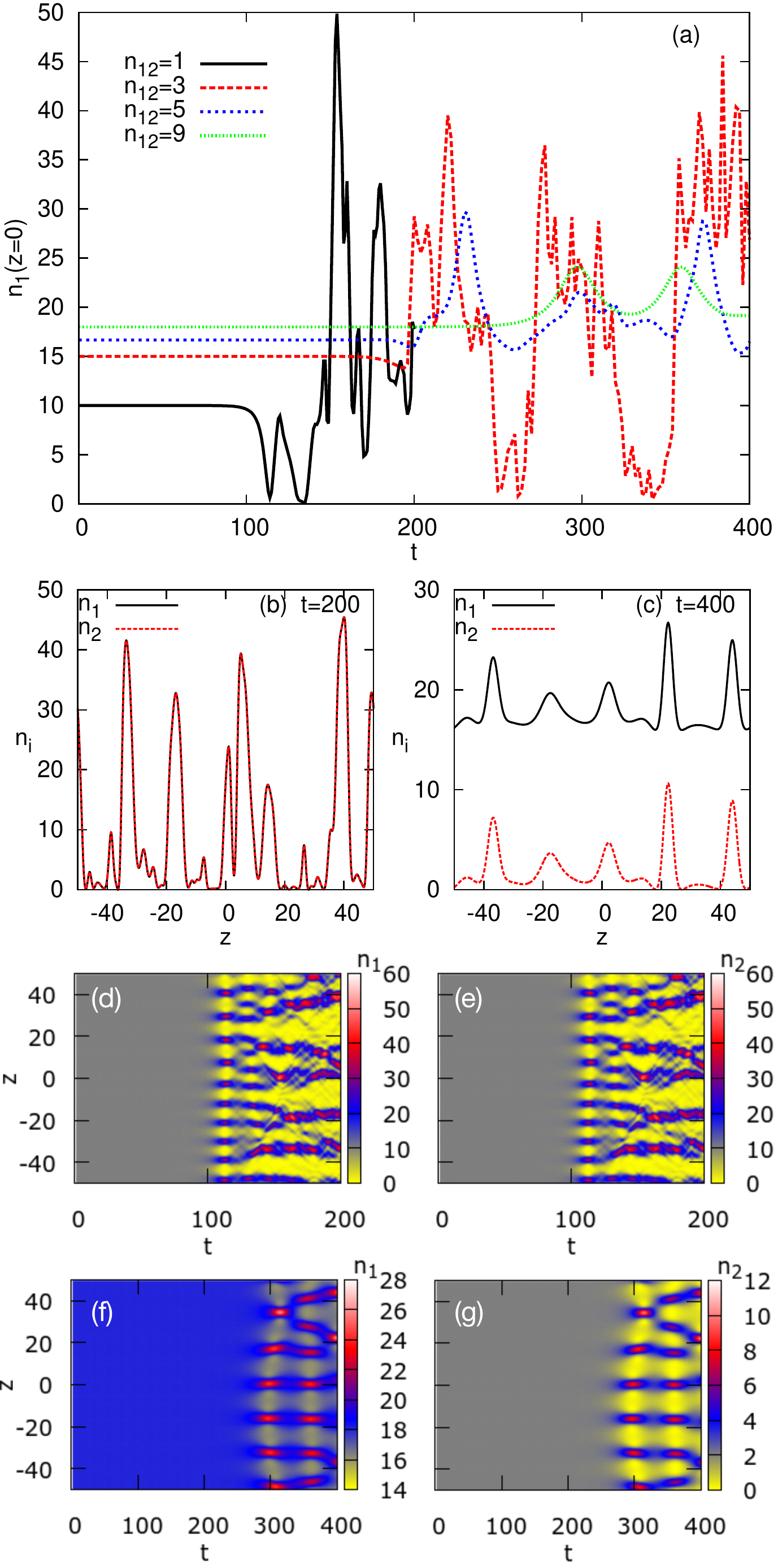}
\caption{Numerically simulated development of the MI of asymmetric PW states
in the two-component system, with $P=1$ and $\protect\delta g/g=0.05$ . The
initial PW states are taken with fixed total density, $(n_{1}+n_{2})/2=10$.
(a) The evolution of the central density of the first component, $n_{1}(z=0)$%
, for different density ratios in the two components, $n_{12}=n_{1}/n_{2}$.
(b,c) Snapshots of density profiles for the cases of (b) $n_{12}\equiv
n_{1}/n_{2}=1$ at $t=200$ and (c) $n_{12}=9$ at $t=400$. Panels (d,e) and
(f,g) are top views of the spatiotemporal evolution of the densities, $%
n_{1}\left( z,t\right) $ and $n_{2}(z,t)$, for $n_{12}=1$ and $n_{12}=9$,
respectively. Simulations were performed in the domain $-50\leq z\leq +50$
with $2048$ grid points, subject to periodic boundary conditions. In this
figure and in Fig.~\protect\ref{homodyn2b}, the scaled time unit corresponds
to $\sim 1$ $\mu$s in physical units.}
\label{homodyn2}
\end{figure}

In Fig.~\ref{homodyn2} we display typical examples of the numerically
simulated development of the MI in the symmetric two-component system with $%
P=1$ and population imbalance. Figure~\ref{homodyn2}(a) shows the evolution
of central-point values of the density of the first component, $n_{1}(z=0)$,
for different values of the density ratio, $n_{12}=n_{1}/n_{2}$. Time
required for the actual onset of the instability increases with the increase
in $n_{12}$, as is clearly shown by the density-plot evolution in Figs.~\ref%
{homodyn2}(d,e) for $n_{12}=1$ and (f,g) for $n_{12}=9$. This observation
can be understood in terms of the MI gain $\sigma $, as shown in Fig.~\ref%
{FigMItwo}(c), where $\sigma $ at $k=k_{\text{max}}$ becomes smaller with
increasing $n_{12}$.

Spatial profiles at fixed time, which are plotted in Fig.~\ref{homodyn2}%
(b,c) for these two cases, show fragmentation of the profiles into sets of
localized structures. The decrease in the number of fragments with the
increase of $n_{12}$ is explained by the decrease of $k_{\max }$, see Fig.~%
\ref{FigMItwo}(a). For $n_{12}=1$, the results are the same as in the
single-component case, as coinciding profiles in the two components of the
symmetric system are stable against spontaneous symmetry breaking. On the
other hand, when $n_{12}\neq 1$ an in-phase two-component localized
structure appears, keeping the initial density imbalance. Since one can
select an arbitrary ratio of densities of the two components for the initial
PW state, a highly asymmetric structure, like the one displayed in Fig.~\ref%
{homodyn2}(c), may emerge even for $P=1$, as a result of the MI-induced
nonlinear evolution. 

\subsubsection{The MI for $P \ne 1$}

\begin{figure}[tbp]
\includegraphics[width=0.95\columnwidth]{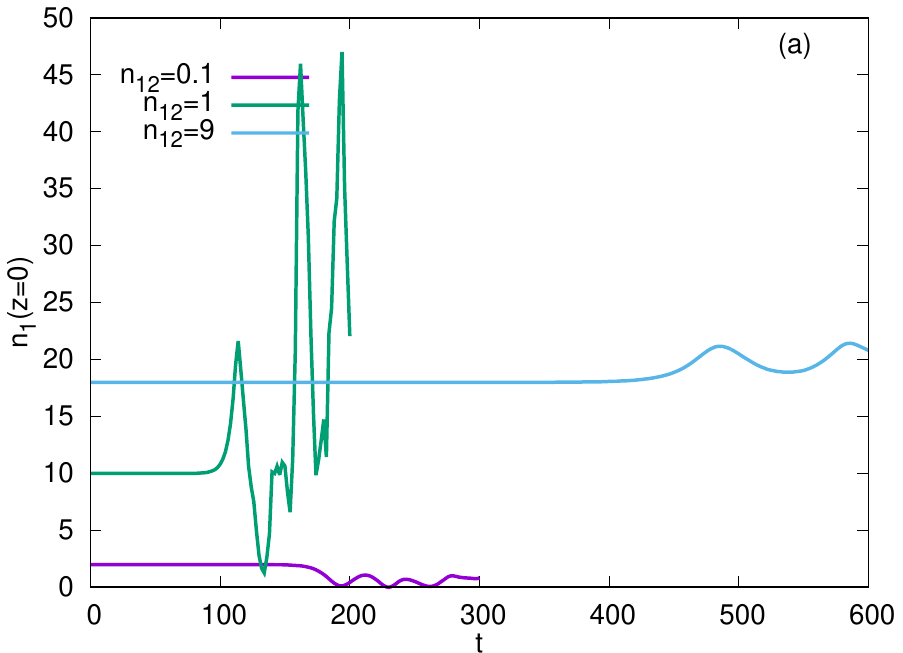}\newline
\includegraphics[width=0.48\columnwidth]{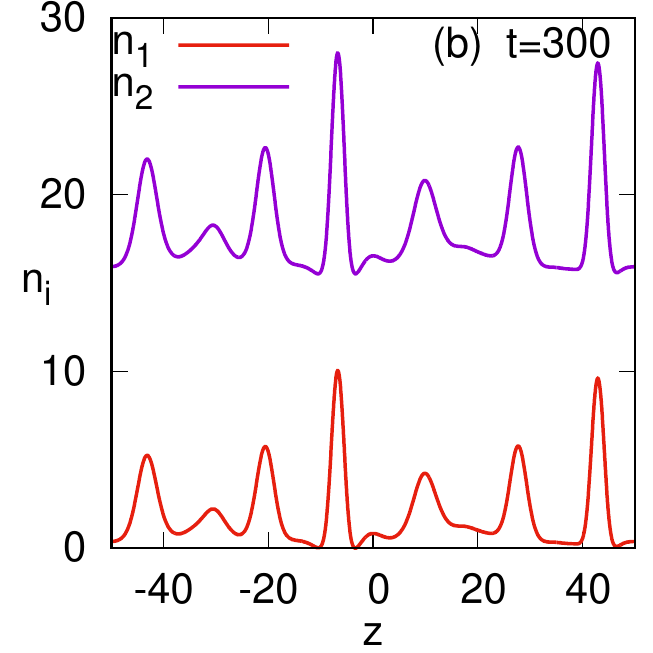}
\includegraphics[width=0.48\columnwidth]{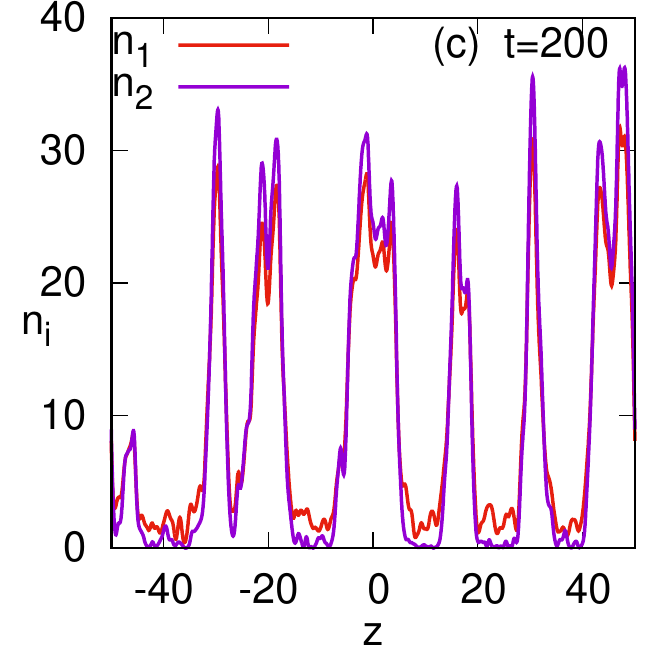}
\includegraphics[width=0.48\columnwidth]{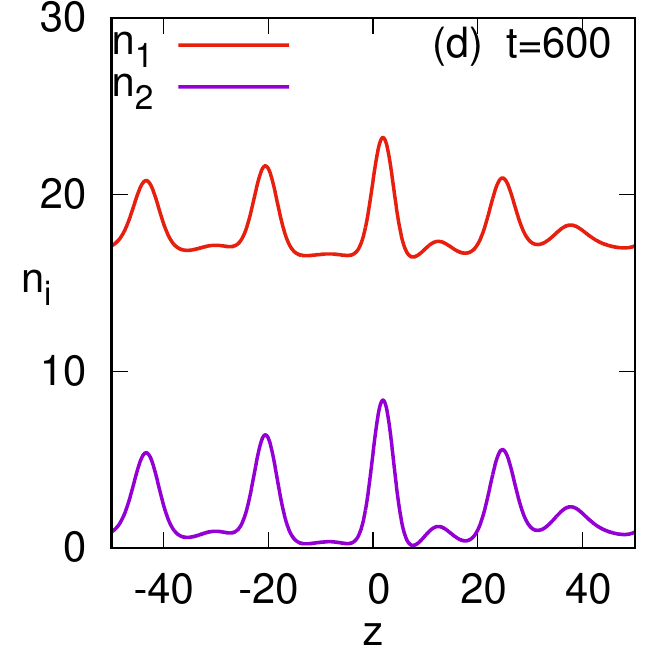} 
\includegraphics[width=0.48\columnwidth]{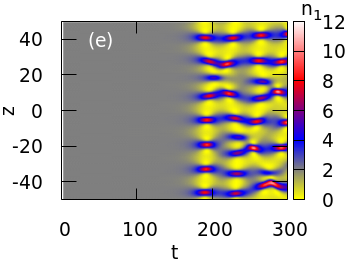} 
\includegraphics[width=0.48\columnwidth]{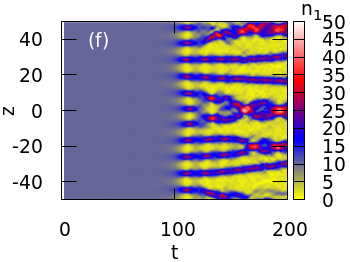} 
\includegraphics[width=0.48\columnwidth]{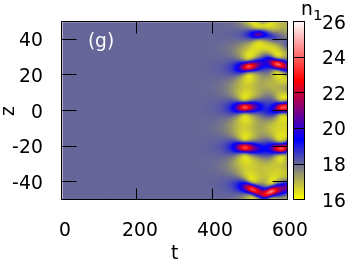}
\caption{Numerically simulated development of the modulational instability
in the two-component system with $\protect\delta g/g=0.05$ and $P=1.25$. The
initial PW states are taken with a fixed total density, $(n_{1}+n_{2})/2=10$%
. (a) The evolution of the central density of the first component, $%
n_{1}(z=0)$, for different density ratios in the two components, $%
n_{12}=n_{1}/n_{2}$. (b-d) Snapshots of density profiles for the cases of
(b) $n_{12}\equiv n_{1}/n_{2}\sim 0.1$ at $t=300$, (c) $n_{12}=1$ at $t=200$
and (d) $n_{12}=9$ at $t=600$. Panels (e-g) represent the top view of the
spatiotemporal evolution of the densities, $n_{1}\left( z,t\right) $,
corresponding to (b-d), respectively (the evolution of $n_{2}\left(
z,t\right) $ shows similar patterns). Simulations were performed in the
domain $-50\leq z\leq +50$ with $2048$ grid points, subject to periodic
boundary conditions. }
\label{homodyn2b}
\end{figure}

Figure~\ref{FigMItwo}(b) represents the MI gain for $P=1.25$ and a fixed
total density, $(n_{1}+n_{2})/2=10$, in the case of slightly different
strengths of the intra-component repulsion. The peak value of the MI\ gain
is attained at $n_{12}=n_{12}^\text{max}=0.577$, below the equal-densities point $n_{12}=1$.
This is consistent with the fact that, at $P>1$, unequal values $n_{1}<n_{2}$
are suitable to the formation of an asymmetric soliton structure, as seen in
Fig.~\ref{Fig_difcritical2}(a). In Fig.~\ref{FigMItwo}(c), we plot the peak
MI\ gain, $\sigma (k_{\text{max}})$, along with the respective value of the
density ratio, $n_{12}^{\text{max}}$, as a function of $P$. Value $n_{12}^{%
\text{max}}$ monotonously decreases as a function of $P$, while the peak
gain attains a minimum at $P=1$.

In Fig.~\ref{homodyn2b}, we present the development of the MI in the
two-component system for $P=1.25$ and a fixed total density, $%
(n_{1}+n_{2})/2=10$. Figure~\ref{homodyn2b}(a) displays the evolution of the
central-point density of the first component, $n_{1}(z=0)$, for different
values of the density ratio, $n_{12}=n_{1}/n_{2}$. It shows that time
required for the development of the MI increases with the increase in the
asymmetry of the density. This is also made evident by the density plots of
the temporal evolution of the first component, shown in\ Figs.~\ref%
{homodyn2b}(e-g). This result is consistent with Eq.~\eqref{MI2gain}, which
shows a decrease of the MI gain with the increase of the asymmetry even for $%
P\neq 1$. Spatial profiles at fixed time, displayed in Fig.~\ref{homodyn2b}%
(b-d), show fragmentation of the profiles. Figure~\ref{homodyn2b}(c) clearly
indicates that, even for $n_{12}=1$, the MI generates asymmetric
droplet-like structures similar to Fig.~\ref{Fig_difcritical2}(a), where the
complete overlapping of the two densities does not occur.

\section{Conclusion}

\label{sec5}  The main purpose of this work is to associate the
MI (modulation instability) of plane waves (PWs) to the mechanism of the
creation of QDs (quantum droplets) in the system described by the coupled GP
(Gross-Pitaevskii) equations including the LHY\ (Lee-Huang-Yang) terms in
the 1D setting. This system is the model of weakly interacting binary Bose
gases with approximately balanced interactions between the intra-component
self-repulsion and the inter-component attraction. We have investigated,
analytically and numerically, the MI of the lower branch of PW states in
both symmetric (effectively single-component) and asymmetric (two-component)
GP systems, and ensuing formation of a chain of droplet-like states. In
particular, numerical solution for QDs which are asymmetric with respect to
the two components are obtained, both in the system with equal repulsion
strengths but unequal populations in the two components, and in the one with
different self-repulsion strengths. The results corroborate that the
previously known symmetric states are robust against symmetry-breaking
disturbances.

These predictions can be tested experimentally by preparing uniform binary
Bose gases with equal or different densities of two components, and suddenly
reducing the strength of the effective MF (mean-field) interaction by means
of the Feshbach-resonance quench, in order to enhance the relative strength
of the LHY terms. In particular, for typical values of physical parameters,
an estimate of the characteristic time of the modulation instability growth
for typical values of the physical parameters is $\sim 1$ $\mu$s.
This time is much smaller than a typical lifetime of the droplet, which is $%
\gtrsim 100$ $\mu$s \cite{cabrera2018quantum}-\cite%
{Semeghini:2018Self}, \cite{long-lived}, thus making the observation of the
MI feasible.\ The present analysis being restricted to the 1D setting,
effects of the tight transverse confinement and crossover to the 3D
configuration \cite{crossover,Ilg:2018,Edler:2018} deserves further
consideration.

\acknowledgments{We appreciate valuable comments received from M. Modugno.
T.M. acknowledges support from IBS (Project Code IBS-R024-D1).
	A.M. acknowledges support from the Ministry of Education, Science and
	Technological Development of Republic of Serbia (project III45010)  and the COST Action CA 16221. The work
	of K.K. is partly supported by the Japan Society for the Promotion of
	Science (JSPS) Grant-in-Aid for Scientific Research (KAKENHI Grant No.
	18K03472). B.A.M. appreciates support from the Israel Science Foundation
	through grant No. 1287/17. A.K. thanks the Indian National Science Academy
	for the grant of INSA Scientist Position at Physics Department, Savitribai
	Phule Pune University.).}



\appendix

\section{Other exact solutions for the single-component GP equation}

\label{othersol1}

Here we briefly list other types of exact solutions of the single-component
equation \eqref{eq13}, in addition to the FT and PW solutions %
\eqref{eq:exact} and \eqref{sing-pert} which were considered in detail above
(solutions to Eq. \eqref{eq13} in the form of dark and anti-dark solitons
were reported in Ref. \cite{Milivoj}).The stability of a majority of these
solutions is not addressed here, as it should be a subject for a separate
work.

\subsection{$\protect\delta g/g > 0$}

In the case of comparable quadratic self-attraction and cubic repulsion in
Eq.~\eqref{eq13} with $\delta g>0$, exact spatially-periodic solutions with
odd parity can be expressed in terms of the Jacobi's elliptic sine, whose
modulus $q$ is an intrinsic parameter of the family:

\begin{equation}
\psi (z,t)=\exp \left( -i\mu _{\text{sn}}t\right) [A~\text{sn}(\beta z,q)+B],
\label{a8}
\end{equation}%
where

\begin{align}
B &=\frac{\sqrt{2}}{3\pi }\frac{g}{\delta g}>0, \quad A=\sqrt{\frac{2}{1+q^{2}%
}}B>0, \\ \quad \mu _{\text{sn}}&=-2\frac{\delta g}{g}B^{2}\,<0,\quad \beta ^{2}=%
\frac{2}{ \left( 1+q^{2}\right) }\frac{\delta g}{g}B^{2}.  \notag
\end{align}%
In the limit of $q\rightarrow 1$, solution (\ref{a8}) goes over into the
kink (the same as found in Ref. \cite{Milivoj}),

\begin{equation}
\psi (z,t)=\exp \left( -i\mu _{\mathrm{kink}}t\right) [A\tanh (\beta z)+B]\,,
\label{a10}
\end{equation}%
with parameters

\begin{equation}
A=B=\frac{\sqrt{2}}{3\pi }\frac{g}{\delta g}>0,\quad \mu _{\mathrm{kink}}=-2%
\frac{\delta g}{g}B^{2},\quad \beta ^{2}=\frac{\delta g}{g}B^{2}.  \notag
\end{equation}

\subsection{$\protect\delta g/g<0$}

In the case when the inter-species MF attraction is stronger than the
intra-species repulsion, resulting in $\delta g<0$, spatially-periodic
solutions are expressed in terms of even Jacobi's elliptic functions, $%
\mathrm{dn}(x,q)$ and $\mathrm{cn}(x,q)$. First, it is

\begin{equation}
\psi (z,t)=\exp \left( -i\mu _{\text{dn}}t\right) [A{\ }\text{dn}(\beta
z,q)+B],  \label{a2}
\end{equation}%
with the elliptic modulus taking all values $0<q<1$, other parameters being

\begin{align}
B &=\frac{\sqrt{2}}{3\pi }\frac{g}{\delta g}<0, \quad A=-\sqrt{\frac{2}{%
2-q^{2}}}B>0, \\ \quad \mu _{\mathrm{dn}} &=-2B^{2}\frac{\delta g}{g}>0, \quad
\beta ^{2}=-\frac{2}{\left(2-q^{2}\right) }\frac{\delta g}{g}B^{2}.  \notag
\label{3}
\end{align}%
The second solution is expressed in terms of the elliptic cosine, with $%
q^{2}>1/2$:

\begin{equation}
\psi (z,t)=\exp \left( -i\mu _{\text{cn}}t\right) [A{\ }\text{cn}(\beta
z,q)+B]\,,  \label{a4}
\end{equation}

\begin{align}
B &=\frac{\sqrt{2}}{3\pi }\frac{g}{\delta g}<0,\quad A=-\sqrt{\frac{2}{%
2q^{2}-1}}B>0,\\ \quad \mu _{cn} &=-2\frac{\delta g}{g}B^{2}\,>0,\quad \beta
^{2}=-\frac{2}{ (2q^{2}-1)}\frac{\delta g}{g}B^{2}.  \notag
\end{align}

In the limit of $q\rightarrow 1$, both solutions (\ref{a2}) and (\ref{a4})
carry over into a state of the ``bubble" type \cite{barashenkov1993stability}%
, which changes the sign at two points (the same solution was reported as an
``W-shaped soliton" in Ref. \cite{Milivoj}):

\begin{equation}
\psi (z,t)=\exp \left( -i\mu _{\mathrm{bubble}}t\right) [A\text{sech}(\beta
z)+B],  \label{6}
\end{equation}%
with parameters
\begin{eqnarray}
B =\frac{\sqrt{2}}{3\pi }\frac{g}{\delta g}<0,\quad A=-\sqrt{2}B\,>0,\\
\quad
\mu _{\mathrm{\ bubble}} = \beta ^{2}=-2\frac{\delta g}{g}B^{2}>0.  \notag
\end{eqnarray}

\section{Analytical solutions for strongly asymmetric fundamental and dipole
states}

Here we consider analytical solutions of Eqs.~(\ref{2comp_GP}) with strong
asymmetry, $N_{1}\ll N_{2},$ which can be found under small-amplitude
conditions, $n_{1}(z=0)\ll n_{2}(z=0)\ll n_{0}$. Then, cubic terms may be
neglected in Eqs.~(\ref{2comp_GP}), and approximation $\sqrt{P|\psi
_{1}|^{2}+P^{-1}|\psi _{2}|^{2}}\approx P^{-1/2}\left\vert \psi
_{2}\right\vert $ is used to simplify Eq.~(\ref{2comp_GP}) to the following
equations for stationary states (\ref{mu}):

\begin{align}
\mu _{1}\phi _{1}& =-\frac{1}{2}\frac{d^{2}\phi _{1}}{dz^{2}}-\frac{\sqrt{P}%
}{\pi }\phi _{2}\phi _{1},  \label{phi1} \\
\mu _{2}\phi _{2}& =-\frac{1}{2}\frac{d^{2}\phi _{2}}{dz^{2}}-\frac{1}{\pi
P^{3/2}}\phi _{2}^{2}.  \label{phi2}
\end{align}%
Although this case is somewhat formal, in terms of the underlying concept of
the quantum droplets, which is essentially based on the competition of
residual MF and LHY\ terms, it is interesting to consider it too.

The soliton solution of Eq.~(\ref{phi2}) is obvious,

\begin{equation}
\phi _{2}(z)=\frac{3\pi }{2}\left( -\mu _{2}\right) \frac{P^{3/2}}{\cosh
^{2}\left( \sqrt{-\mu _{2}/2}z\right) }  \label{sech^2}
\end{equation}%
[solution (\ref{eq:exact}) takes essentially the same form in the limit of $%
|\mu |\ll \mu _{0}$]. Then, the substitution of Eq.~(\ref{sech^2}) in Eq.~(%
\ref{phi1}) makes it tantamount to the linear Schr\"{o}dinger equation with
the\ P\"{o}schl-Teller potential \cite{LL}. The ground-state (GS) solution
of Eq.~(\ref{phi1}) for $\phi _{1}$, with arbitrary amplitude $\phi
_{1}^{(0)}$,

\begin{equation}
\left( \phi _{1}(z)\right) _{\mathrm{GS}}=\frac{\phi _{1}^{(0)}}{\left[
\cosh \left( \sqrt{-\mu _{2}/2}z\right) \right] ^{\gamma }},  \label{GS}
\end{equation}%
exists with

\begin{equation}
\gamma =\frac{1}{2}\left( \sqrt{24P^{2}+1}-1\right) ,  \label{gamma-GS}
\end{equation}%
and eigenvalue

\begin{equation}
\left( \mu _{1}\right) _{\mathrm{GS}}=\left( \sqrt{24P^{2}+1}-1\right) ^{2}\
\frac{\mu _{2}}{16}.  \label{muGS}
\end{equation}%
In this case, the QD solutions are quasi-Gaussian objects \cite%
{Astrakharchik:2018Dynamics}. Note that, in the symmetric system with $P=1$,
Eqs. (\ref{gamma-GS}) and (\ref{muGS}) yield $\gamma =2$ and $\left( \mu
_{1}\right) _{\mathrm{GS}}=\mu _{2}$, i.e., the eigenmode and eigenvalue
coincide with their counterparts in the soliton solution (\ref{sech^2}),
while they are different in the asymmetric system, the GS level lying below
or above the chemical potential of soliton (\ref{sech^2}) at $g_{1}>g_{2}$
and $g_{1}<g_{2}$, respectively.

\begin{figure}[tbh]
\includegraphics[width=6.5cm,scale=1]{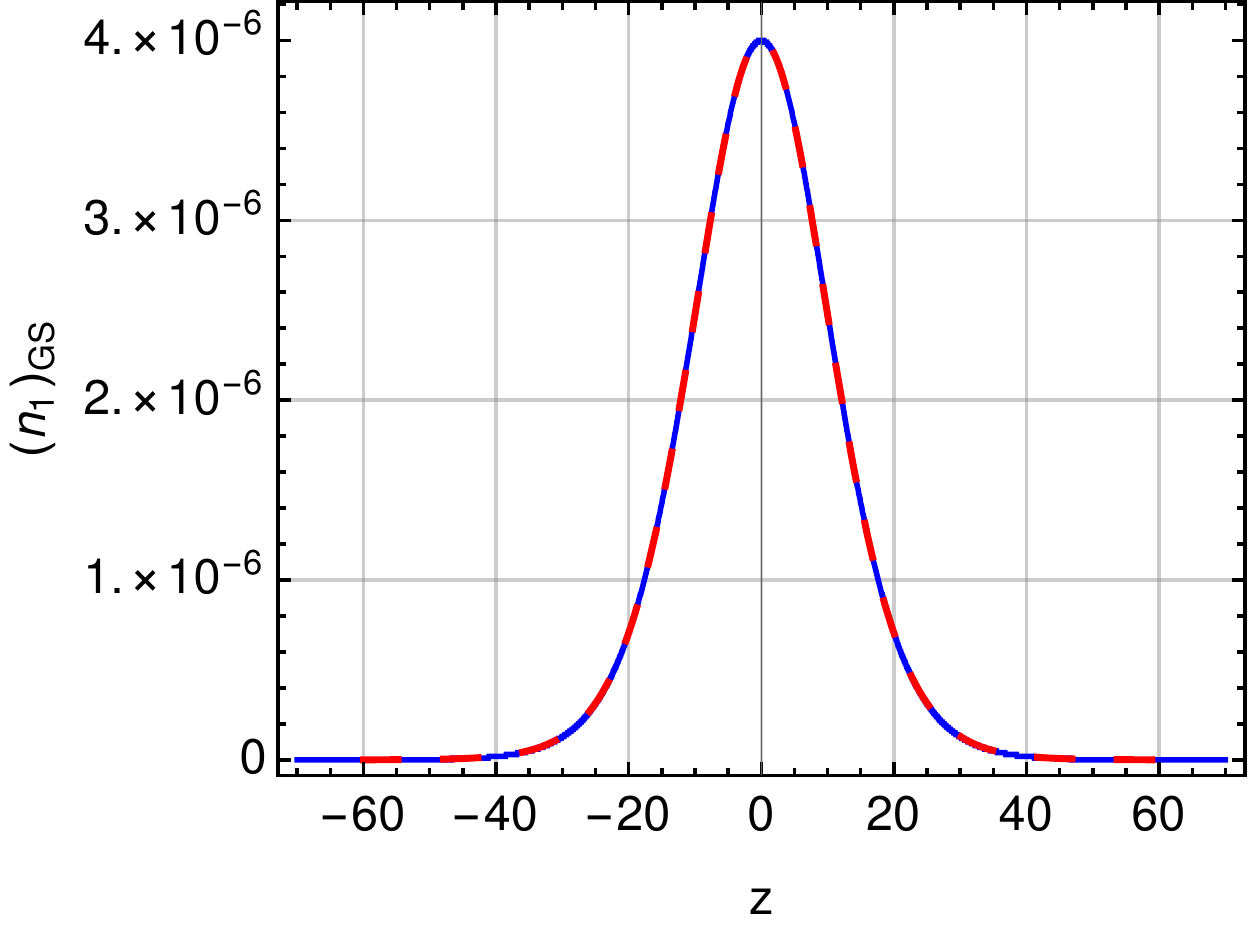} %
\includegraphics[width=6.5cm,scale=1]{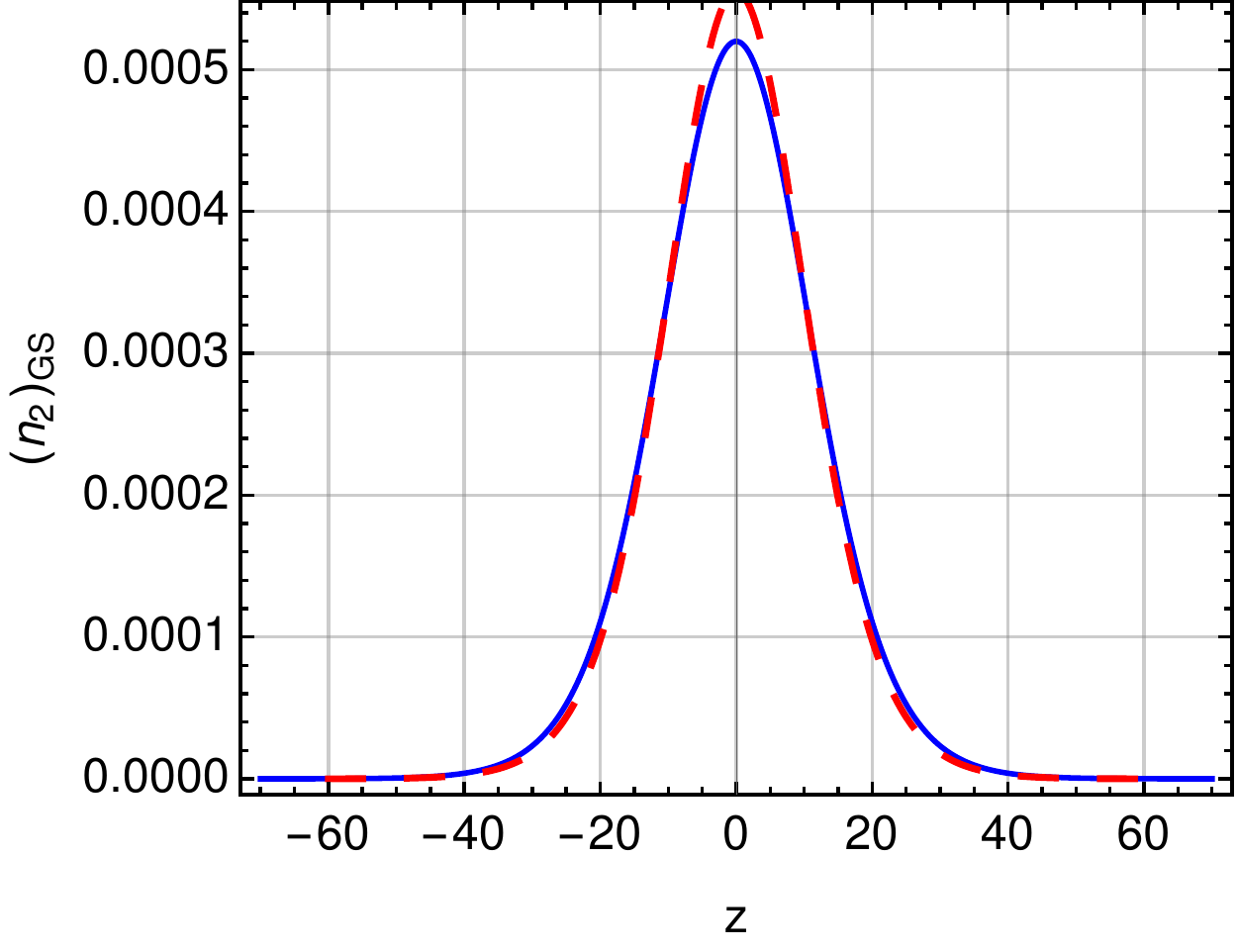}
\caption{Comparison of the asymptotic analytical solutions, given by Eqs.~
\eqref{sech^2} and \eqref{GS}, with their numerically obtained counterparts.
The density of the first ($n_{1}$) and second ($n_{2}$) components are
displayed in top and bottom panels, respectively. Solid blue lines represent
the numerical results, while dashed red lines represent the analytical
solution. Here, parameters are $\protect\delta g/g=0.05$, $N_{1}=0.0001067$,
$N_{2}=0.0148044$ and $(\protect\mu _{2})_{\mathrm{GS}}=\protect\mu %
_{2}=-0.005$. }
\label{fig:ns_as}
\end{figure}

In Fig.~\ref{fig:ns_as} we compare a typical asymptotic solution given by
Eqs.~\eqref{sech^2} and \eqref{GS} with a numerically obtained GS solution
for the same values of the parameters. It is seen that the analytical and
numerical results match well.

Further, it is also possible to produce the first excited state of Eq.~(\ref%
{phi1}) in the form of the dipole (antisymmetric) mode with an arbitrary
amplitude:

\begin{equation}
\left( \phi _{1}(z)\right) _{\mathrm{dip}}=\frac{\phi _{1}^{(0)}\sinh \left(
\sqrt{-\mu _{2}/2}z\right) }{\left[ \cosh \left( \sqrt{-\mu _{2}/2}z\right) %
\right] ^{\gamma }},  \label{dip}
\end{equation}%
where $\gamma $ is the same as in Eq.~(\ref{gamma-GS}), the respective
eigenvalue being

\begin{equation}
\left( \mu _{1}\right) _{\mathrm{dip}}=\left( \sqrt{24P^{2}+1}-3\right) ^{2}%
\frac{\mu _{2}}{16},  \label{mudip}
\end{equation}%
which is obviously higher than its GS counterpart (\ref{muGS}) [at $P=1$,
Eq.~(\ref{mudip}) yields $\left( \mu _{1}\right) _{\mathrm{dip}}=\mu _{2}/4$%
, and $\left( \mu _{1}\right) _{\mathrm{dip}}$ falls below $\mu _{2}$ for $P>%
\sqrt{2}$]. Unlike the GS, the dipole mode exists not at all values of $P$,
but only for $P>\sqrt{1/3}$. Exactly at $P=\sqrt{1/3}$, one has $\left( \mu
_{1}\right) _{\mathrm{dip}}=0$, and the dipole mode (\ref{dip}), with $%
\gamma =1$, is a delocalized one, $\sim \tanh \left( \sqrt{-\mu _{2}/2}%
z\right) $.

Linear Schr\"{o}dinger equation (\ref{phi1}) with the P\"{o}schl-Teller
potential may give rise to higher bound states of integer order $\nu $ as
well, with eigenvalues

\begin{equation}
\left( \mu _{1}\right) _{\nu }=\left( \sqrt{24P^{2}+1}-\left( 1+2\nu \right)
\right) ^{2}\frac{\mu _{2}}{16},  \label{nu}
\end{equation}%
where $\nu =0$ and $1$ correspond to Eqs.~(\ref{muGS}) and (\ref{mudip}),
respectively, the $\nu $-th spate existing at $P^{2}>\nu \left( \nu
+1\right) /6$. The number of such solutions is always finite.

Unlike solutions considered in Appendices A and C, the stability of
solutions given by Eqs. (\ref{sech^2})-(\ref{nu}) is obvious.

\section{Other exact solutions in the case of $N_{1}\ll N_{2}$}

\label{othersol2} Here we provide periodic solutions to the semi-linear
system of Eqs.~\eqref{phi1} and \eqref{phi2} in terms of Jacobi elliptic
functions. In the limit of $q\rightarrow 1$, they go over into solutions
given in the main text, in the form of Eqs.~\eqref{sech^2}, \eqref{GS} and %
\eqref{dip}.

\subsection{ Solution of Eq.~(\protect\ref{phi2})}

An exact periodic solution of Eq.~(\ref{phi2}) with the quadratic
nonlinearity is

\begin{equation}
\phi _{2}=A[\text{dn}^{2}(\beta z,q)+p]\,,  \label{3app}
\end{equation}%
with

\begin{equation}
\begin{split}
\beta^{2} &=-\frac{\mu _{2}}{2\sqrt{1-q+q^{2}}}, \quad A=-\frac{3\pi \mu
_{2}P^{3/2}}{2\sqrt{1-q+q^{2}}},\\ \quad p &=\frac{-(2-q)+\sqrt{1-q+q^{2}}}{3}.
\label{4}
\end{split}
\end{equation}%
In the limit of $q\rightarrow 1$, solution (\ref{3app}) goes over into
solution \eqref{sech^2}. Note that $p$ is vanishing in this limit, according
to Eq. (\ref{4}).

\subsection{Solutions of Eq.~(\protect\ref{phi1})}

We now show that, with $\phi _{2}$ given by Eq.~(\ref{3app}), linear
equation (\ref{phi1}) $\phi _{1}$ has several particular solutions depending
on the value of $P$.

\textbf{Solutions For $P^2 = 1/3$}

\subsubsection{Solution I}

It is easy to check that

\begin{equation}
\phi _{1}=\phi _{1}^{(0)}\text{dn}(\beta z,q)  \label{5app}
\end{equation}%
is an exact solution to Eq.~(\ref{phi1}), provided that

\begin{equation}
P^{2}=\frac{1}{3}, \quad \mu _{1}=\left( \frac{\mu _{2}}{12}\right) \frac{%
2-q+2 \sqrt{1-q+q^{2}}}{\sqrt{1-q+q^{2}}}\,.  \notag
\end{equation}

\subsubsection{ Solution II}

\begin{equation}
\phi _{1}=\phi _{1}^{(0)}\text{cn}(\beta z,q)  \label{7app}
\end{equation}%
is an exact solution to Eq.~(\ref{phi1}), provided that

\begin{equation}
P^{2}=\frac{1}{3},\quad \mu _{1}=\left( \frac{\mu _{2}}{12}\right) \frac{%
2q-1+2 \sqrt{1-q+q^{2}}}{\sqrt{1-q+q^{2}}}\,.  \notag
\end{equation}%
In the limit of $q\rightarrow 1$, solutions I and II go over into the
solution Eq.~\eqref{GS} with $\gamma =1$ and $\mu _{1}=\mu _{2}/4$.

\subsubsection{ Solution III}

\begin{equation}
\phi _{1}=\phi _{1}^{(0)}\text{sn}(\beta z,q)  \label{9}
\end{equation}%
is an exact solution to Eq.~(\ref{phi1}), provided that

\begin{equation}
P^{2}=\frac{1}{3}\,,\quad \mu _{1}=\left( \frac{\mu _{2}}{12}\right) \frac{2%
\sqrt{ 1-q+q^{2}}-(1+q)}{\sqrt{1-q+q^{2}}}\,.  \notag
\end{equation}%
In the limit of $q\rightarrow 1$, solution III goes over into the solution
Eq.~\eqref{dip} with $\gamma =1$ and $\mu _{1}=0$.

\textbf{Solutions For $P^2 = 1$}

\subsubsection{ Solution IV}

It is easy to check that

\begin{equation}
\phi _{1}=\phi _{1}^{(0)}[\text{dn}^{2}(\beta z,q)+p]\,  \label{11}
\end{equation}%
is an exact solution to Eq.~(\ref{phi1}), provided that

\begin{equation}
P^{2}=1, \quad \mu _{1}=\mu _{2}\,.  \notag
\end{equation}

\subsubsection{ Solution V}

\begin{equation}
\phi _{1}=\phi _{1}^{(0)}\text{cn}(\beta z,q)\text{dn}(\beta z,q)\,
\label{13}
\end{equation}%
is an exact solution to Eq.~(\ref{phi1}), provided that

\begin{equation}
P^{2}=1,\quad \mu _{1}=\left( \frac{\mu _{2}}{2}\right) \frac{q+\sqrt{%
1-q+q^{2}}}{\sqrt{1-q+q^{2}}}\,.  \notag
\end{equation}%
In the limit $q=1$, solutions IV and V go over into solution Eq.~\eqref{GS}
with $\gamma =2$ and $\mu _{1}=\mu _{2}$.

\subsubsection{ Solution VI}

\begin{equation}
\phi _{1}=\phi _{1}^{(0)}\text{sn}(\beta z,q)\text{dn}(\beta z,q)\,
\label{15}
\end{equation}%
is an exact solution to Eq.~(\ref{phi1}), provided that

\begin{equation}
P^{2}=1\,,\quad \mu _{1}=\left( \frac{\mu _{2}}{4}\right) \frac{3(1-q)+\sqrt{
1-q+q^{2}}}{\sqrt{1-q+q^{2}}}\,.  \notag
\end{equation}

\subsubsection{ Solution VII}

\begin{equation}
\phi _{1}=\phi _{1}^{(0)}\text{sn}(\beta z,q)\text{cn}(\beta z,q)\,
\label{17}
\end{equation}%
is an exact solution to Eq.~(\ref{phi1}), provided that

\begin{equation}
P^{2}=1\,,\quad \mu _{1}=\left( \frac{\mu _{2}}{4}\right) \frac{2\sqrt{%
1-q+q^{2}} -(2-q)}{\sqrt{1-q+q^{2}}}\,.  \notag
\end{equation}%
In the limit of $q\rightarrow 1$, solutions VI and VII go over into %
\eqref{dip}, with $\gamma =2$ and $\mu _{1}=\mu _{2}/4$.



\end{document}